\documentclass[preprint,10pt]{elsarticle}

\usepackage[margin=0.8in]{geometry}
\usepackage{amsmath,amssymb,amsthm}
\usepackage{mathtools} 
\usepackage{dsfont}
\usepackage{stmaryrd}
\usepackage{float}
\usepackage{subfigure}
\usepackage{graphicx}
\usepackage{xcolor}
\usepackage{tabu}
\usepackage{multicol}
\usepackage{lipsum}	
\usepackage{caption}
\usepackage{longtable}
\usepackage{booktabs}

\newcommand{\nes}{\hspace*{-0.6pt}}
\newcommand{\exs}{\hspace*{0.6pt}}

\newcommand{\bff}{\boldsymbol{f}}

\newcommand{\bfe} {\boldsymbol{e}}

\newcommand{\GamN}{\Gamma_{\!\mbox{\tiny{N}}}}
\newcommand{\GamD}{\Gamma_{\!\mbox{\tiny{D}}}}
\newcommand{\Gamobs}{\Gamma_{\!\mathrm{obs}}}
\newcommand{\bfeps}{\boldsymbol{\varepsilon}}

\newcommand{\bfalpha}{\boldsymbol{\alpha}}
\newcommand{\bfbeta}{\boldsymbol{\beta}}
\newcommand{\bfeta}{\boldsymbol{\eta}}
\newcommand{\bfxi}{\boldsymbol{\xi}}

\newcommand{\bfuobs}{\boldsymbol{u}^{\mathrm{obs}}}
\newcommand{\bfI} {\boldsymbol{I}}
\newcommand{\bfn} {\boldsymbol{n}}

\newcommand{\bfM} {\boldsymbol{M}}
\newcommand{\bfH} {\boldsymbol{H}}
\newcommand{\bfA} {\boldsymbol{A}}
\newcommand{\bfB} {\boldsymbol{B}}
\newcommand{\bfC} {\boldsymbol{C}}
\newcommand{\bfv} {\boldsymbol{v}}
\newcommand{\bfvv} {\textbf{v}}
\newcommand{\bfu} {\boldsymbol{u}}
\newcommand{\bfg} {\boldsymbol{g}}
\newcommand{\bfh} {\boldsymbol{h}}

\newcommand{\bfsig}{\boldsymbol{\sigma}}

\newcommand{\dd}{\,\text{d}}

\newcommand{\bfx} {\boldsymbol{x}}

\newcommand{\bfze} {\mathbf{0}}

\newcommand{\dip} {\! :\!}

\newcommand{\IS}{\mbox{\boldmath $\mathcal{I}$}}

\newcommand{\rec}{_{\text{rec}}}

\renewcommand{\Re}{\text{Re}}
\renewcommand{\Im}{\text{Im}}

\newcommand{\hh}{\hspace*{0.7pt}}

\newcommand{\sip} {\! \cdot \!}

\newcommand{\tdemi} {\tfrac{1}{2}\exs}

\usepackage[algoruled,vlined,boxed,longend,english,lined,boxed,commentsnumbered]{algorithm2e} 	
\usepackage{algorithmic}

\newtheorem{remark}{Remark}

\definecolor{orange-red}{rgb}{1.0, 0.5, 0.0}

\begin{document}
\begin{frontmatter}

\title{Full-waveform reconstruction of micro-seismic events via topological derivative approach}

\author[1]{A.A.M. da Silva}
\author[1]{A.A. Novotnty}
\author[2]{A.A.S. Amad}
\author[3]{B.B. Guzina \corref{BG}}

\address[1]{Laborat\'{o}rio Nacional de Computa\c{c}\~{a}o Cient\'{\i}fica LNCC/MCTI, Petr\'{o}polis, Brazil}
\address[2]{College of Engineering, Swansea University Bay Campus, Swansea, Wales, UK}
\address[3]{Department of Civil, Environmental, and Geo-Engineering, University of Minnesota Twin Cities, US}

\cortext[BG]{Corresponding author (guzin001@umn.edu)}

\date{}

\begin{abstract}
Micro-seismic events, naturally occurring within geological formations and quasi-brittle engineered systems, provide a powerful window into the evolving processes of material degradation and failure. Accurate characterization of these events is critical for a comprehensive understanding of the underpinning fracturing mechanisms and potential implications.  In this work, we present an algorithm for the spatial reconstruction and characterization of micro-seismic events in a three-dimensional bounded elastic body (with known geometry and nominal material properties) via joint source location and moment tensor inversion. Assuming  availability of the full-waveform “acoustic emission” traces whose spectral content can be exposed via Fourier transform, the inverse solution is based on (i) a time-harmonic (forward) elastodynamic model and (ii) the concept of topological derivative as a framework for robust event reconstruction. On exploiting an equivalence between the elastic wavefield generated by the creation of a new micro-surface and that stemming from a suitable set of dipoles and double-couples (whose strengths are synthesized via the seismic moment tensor), we formulate the inverse problem as that for the real (in-phase) and imaginary (out-of-phase) components of the moment tensor at trial grid locations. In this way the optimal solution is obtained via a combinatorial search over a prescribed grid, inherently allowing for successive refinements of event reconstruction over the region(s) of interest. The analysis is illustrated by numerical experiments highlighting the key features of the inversion scheme including the reconstruction of multiple (i.e.~contemporaneous) events,  localization of the ``off-grid'' micro-seismic events, and the ability to handle noisy data. The results in particular highlight the utility of multi-frequency event reconstruction toward reducing the demand on the number of sensing locations. 
\end{abstract}


\end{frontmatter}

\section{Introduction} \label{sec:introduction}

\noindent Accurate reconstruction of the source locations and affiliated moment tensors characterizing micro-seismic events is of paramount importance in seismology, geophysics, reservoir engineering, mining operations, maintaining civil infrastructure, and the advancement of geological CO$_2$ storage \citep{KoenerRM1981,manthei2018review,okoli2024alterations,ShearerBook2009}. In particular, such information helps elevate our understanding of the \emph{in situ} stress distribution, fault mechanics, evolution of a (localized or diffuse) failure event, or the progress of reactive flow. On the laboratory scale, acoustic emission sensing is equally essential to the fundamental understanding of fracturing processes in quasi-brittle materials~\cite{labuz2001acoustic,maji1990fracture}. 

Conventional approaches to the interpretation of acoustic emission signals \cite[e.g.][]{carpinteri2012reliable,ono2014acoustic} often provide limited spatial resolution in terms of the event location and lack the ability to fully characterize the source mechanism. The latter can be effectively synthesized via the seismic moment tensor~\citep{AkiBook2009}, which furnishes information about the kinematics of a displacement field discontinuity giving rise to the newly generated micro-crack surface. To overcome the limitations of conventional analyses, the concept of full-waveform inversion (FWI) has emerged as a lynchpin that leverages the entirety of recorded seismic traces in the pursuit of parameter estimation. In short, FWI aims to find the best-fit model that minimizes the discrepancy between observed and synthetic waveforms (i.e. traces) under the constraint of e.g. Navier equations. The inclusion of moment tensor information in the FWI inversion opens the path toward both elevated location accuracy and in-depth understanding of the observed micro-seismic events. 

The essential framework behind reconstructing the seismic moment tensor from the full-waveform observations of a (vector) displacement field at several remote points was established by \citeauthor{GilbertPTRS1973} \cite{GilbertPTRS1973} using the theory of linear elastic wave propagation. In this vein, \citeauthor{RiceJNE1980} \cite{RiceJNE1980} introduced the theory of ``acoustic'' i.e. elastic wave emission from a damage processes due to micro-cracking by describing the elastic wavefield emanating from a damage event in terms of seismic moment tensor densities supported over the source region. The authors in \cite{SjogreenJSC2014} studied the inverse problem of estimating the seismic moment tensor characterizing a point source in the context of FWI, where a non-quadratic (and generally non-convex) misfit functional is minimized via gradient-based optimization. An application of FWI moment tensor reconstruction to hydrofracture monitoring can be found in~\cite{SongG2011}, where the cost functional (quantifying the misfit between observed and synthetic time traces) is minimized via a grid search in the event-origin-time-and-location space. Related field applications targeting the seismic moment tensor include monitoring of micro-seismicity due to CO$_2$ injection \cite{QinEnergies2023}, and analysis of the 2020 El-Negalah earthquake \cite{ElhadidyJAG2023}. More recently, the authors in \cite{AmadIJSS2020} proposed a full-waveform inversion technique for the spatial reconstruction and moment-tensor characterization of two-dimensional (2D) micro-seismic events in the frequency domain. An attempt at simultaneous source and (heterogeneous) medium reconstruction by way of frequency-domain FWI can be found in \cite{SongJSTAEORS2019}; however the authors consider the scalar wave equation which voids the need for reconstructing the seismic moment tensor.  

In this work we focus on the joint source location and moment tensor inversion of micro-seismic events in a 3D finite elastic body, catering for the laboratory investigation of damage and fracture processes in natural and engineered materials. To this end, we assume (sparse) full-waveform, acoustic emission sensing on the boundary of a finite specimen with known geometry, dimensions, and material (elastic) properties. In this setting, we consider relevant frequency-domain components of the recorded time traces, whereby the forward elastodynamic system is governed by the time-harmonic Navier equations. The micro-seismic events are described as combinations of dipoles and double couples~\citep{AkiBook2009}, whose respective amplitudes are compiled by a complex-valued seismic moment tensor. The basic idea consists in rewriting the inverse problem as an optimization problem whose functional (measuring a misfit between the observed and predicted time-harmonic data) is minimized with respect to the set of admissible source characteristics -- leading to a non-iterative, grid-search reconstruction algorithm. The key advantages over the earlier FWI strategies~\cite[e.g.][]{SongG2011} include reduced dimension of the search domain (space vs. space-time) and accelerated computation of the forward problem (frequency- vs. time-domain)  for a finite elastic body. Since any micro-seismic event by definition entails creation of a new infinitesimal fracture surface, the proposed reconstruction algorithm is seen as a specialization of the topological derivative method described for instance in \cite{AmmariBook2004,AmmariBook2013,RivaBook2021,MazyaBook2000},  For an account of the theoretical basis and applications of the latter asymptotic framework, the reader is referred to~\cite{NovotnyBook2013,NovotnyBook2020}.

The paper is organized as follows. In Section \ref{sec:seismic}, we introduce the forward problem and inverse source problem for an elastic body, including the formulation of the non-iterative inverse solution. The latter includes the sensitivity analysis of the misfit functional with respect to source perturbations within the set of admissible solutions, and an algorithm for the reconstruction and characterization of micro-seismic events in a 3D elastic body. Numerical experiments are presented in Section \ref{sec:numerics}, showing the efficiency and accuracy of the reconstruction algorithm. In particular, the results highlight the ability of the inverse solution to handle multiple (i.e.~contemporaneous) events, noisy data, and  accurate localization of the ``off-grid'' micro-seismic events. We also demonstrate that the use of multi-frequency data carries a significant potential  toward reducing the demand on the number of sensing locations.

\section{Seismic moment tensor inversion} \label{sec:seismic}

\noindent Consider a finite elastic body $\Omega\!\subset\!\mathds{R}^{3}$ with Lipschitz boundary $\partial\Omega$ that is endowed with the mass density~$\rho$ and elasticity tensor $\bfC\in\mathds{R}^{3\times 3\times 3\times 3}$. Let~$\GamN \subset \partial \Omega$ and~$\GamD = \partial\Omega\backslash\GamN$ denote the boundary segments subjected to the homogeneous Neumann and homogeneous Dirichlet boundary conditions, respectively. In the sequel, we consider time-harmonic motions with implicit time factor $e^{-i \omega t}$, where $t$ is the time variable, $\omega$ denotes the frequency of wave motion and $i=\sqrt{-1}$ is the imaginary unit. Letting further~$\bfsig$ denote the Cauchy stress tensor in~$\Omega$ whereby 
\begin{equation}
\bfsig[\bfu] = \bfC \dip \bfeps[\bfu] ,
\end{equation}
where the strain tensor $\bfeps$ is given by the symmetric part of the gradient of $\bfu$, namely
\begin{equation}
\bfeps[\bfu] = \tfrac{1}{2} \big(\nabla\bfu +\nabla^{\mbox{\tiny T}}\!\bfu\big) .
\end{equation}
We are interested in the \emph{inverse source problem} of reconstructing the source density~$\widehat{\bff}$, such that 
\begin{equation}
\left\{
\begin{array}{rllcl}
-\nabla\sip \bfsig[\bfu] - \rho\exs\omega^2 \bfu & = &\widehat{\bff} & \text{in} & \Omega , \\
\bfu & = & \bfuobs & \text{on} & \Gamobs\subset \GamN,\\
\bfu & = & \bfze & \mbox{on} & \GamD , \\
\bfn\sip \bfsig[\bfu] & = & \bfze & \text{on} & \GamN,
\end{array}
\right.
\label{eq:strongZ}
\end{equation}
where $\bfu\!\!:\! \Omega \mapsto \mathds{C}^{3}$ is the elastodynamic displacement field; $\bfn$ is the unit outward normal vector on~$\partial{\Omega}$; $\Gamobs\subset\GamN$ is the measurement surface, and~$\bfuobs$ are the (time-harmonic) ``acoustic emission'' data from which we aim to resolve~$\widehat{\bff}$. Hereon, we assume the elastic body~$\Omega$ to be homogeneous and isotropic, in which case the elasticity tensor reads  
\begin{equation}
\bfC = 2 \mu \, \IS + \lambda \, \bfI\! \otimes \!\bfI,
\label{eq:TensorC}
\end{equation}
where~$\lambda$ and~$\mu$ are the Lam\'{e} moduli, while $\bfI$ and $\IS$ denote respectively the second-order and symmetric fourth-order identity tensor. Using the Einstein summation convention over repeated indices $i,j,k,l = \overline{1,3}$, we specifically have $\bfI=\delta_{ij}\,\bfe_i\!\otimes\!\bfe_j$ and $\IS=\tdemi(\delta_{ik}\delta_{jl}+\delta_{il}\delta _{jk})\,\bfe_i\!\otimes\!\bfe_j\!\otimes\!\bfe_k\!\otimes\!\bfe_l$, where  $\bfe_j$ is the unit vector in the (Cartesian) $j$th  coordinate direction. For future reference, we recall the expressions for the Lam\'{e} parameters
\begin{equation}
   \lambda = \frac{\nu E}{(1+\nu)(1-2\nu)} \quad \text{and} \quad \mu=\frac{E}{2(1+\nu)},
\end{equation}
in terms of the Young's modulus $E$ and Poisson's ratio $\nu$.

Motivated by the application to acoustic emission problems, we next describe the source density $\widehat{\bff}$ via the superposition of a finite number of dipoles and double couples~\citep{AkiBook2009}. More precisely, we assume that $\widehat{\bff} \in C_\delta(\Omega)$, where 
\begin{equation} 
C_\delta(\Omega) = \Big\{ \bff \!: \Omega \rightarrow \mathds{C}^{3}\;|\; \bff(\bfx) = \sum_{n=1}^{N} \bfM^{(n)} \sip \nabla_{\!\!\bfx^{(n)}} \delta(\bfx-\bfx^{(n)}), ~ \bfx^{(n)}\!\nes\in\Omega \Big\}
\label{eq:Cdelta}
\end{equation}
for some finite~$N$. Here, $\delta(\boldsymbol{\cdot})$ is the three-dimensional Dirac delta function; $N$ denotes the number of point sources located at $\bfx^{(n)} \in \Omega$, and $\bfM^{(n)} \!\in \mathds{C}^{3\times 3}$ is a (symmetric) \emph{seismic moment tensor} characterizing the~$n$th point source ($n\!=\!\overline{1,N}$). A more detailed structure of~$\bfM^{(n)}$ corresponding to various dipoles and double couples will be discussed shortly. On the basis of~\eqref{eq:Cdelta}, we write the sought source density satisfying~\eqref{eq:strongZ} as
\begin{equation} 
\widehat{\bff}(\bfx) = \sum_{n=1}^{\widehat{N}} \widehat{\bfM}^{(n)} \!\sip\, \nabla_{\!\!\bfx^{(n)}} \delta(\bfx - \widehat{\bfx}^{(n)}).
\label{eq:truef}
\end{equation}
In this setting, our task consists in reconstructing $\widehat{\bff} \in C_\delta(\Omega)$ in terms of the micro-seismic event locations $\widehat{\bfx}^{(n)}$ and moment tensors $\widehat{\bfM}^{(n)}$ ($n=1,2,\ldots\widehat{N}$), from pointwise boundary measurements $\bfuobs$. 

Assuming the inverse problem \eqref{eq:strongZ} to be over-determined, we rewrite it as a constrained optimization problem featuring the least-squares functional
\begin{equation}
\mathcal{J}(\bfu) = \tfrac{1}{2} \int_{\Gamobs} \! \mathfrak{s}(\bfx) \, \|\bfu-\bfuobs\|^2 \dd\Gamma,
\label{eq:funcOri}
\end{equation}
where $\|\bfv\| = \sqrt{\bfv \sip \bfv^\ast}$, with $(\cdot)^\ast$ denoting complex conjugation of $(\cdot)$; 
\begin{equation}\label{fraks}
\mathfrak{s}(\bfx) = \sum_{\mathfrak{m}} \delta(\bfx-\bfx^{\text{obs}}_\mathfrak{m}),
\end{equation}
$\bfx^{\text{obs}}_\mathfrak{m}\in\Gamobs$ are the coordinates of the measurement points, and $\bfu\!: \Omega \mapsto \mathds{C}^{3}$ solves the boundary value problem
\begin{equation}
\left\{
\begin{array}{rllcl}
-\nabla\sip \bfsig[\bfu] - \rho \exs \omega^2 \bfu & = & {\bff} & \text{in} & \Omega, \\
\bfu & = & \bfze & \mbox{on} & \GamD , \\
\bfn \sip \bfsig[\bfu] & = & \bfze & \text{on} & \GamN .
\end{array} \right.
\label{eq:strongOri}
\end{equation}
As a result, the optimization problem at hand can be stated as 
\begin{equation}
\underset{\bff \in C_\delta(\Omega)}{\text{Minimize}} \; \mathcal{J}(\bfu), \; \text{subject to \eqref{eq:strongOri}}.
\label{eq:min}
\end{equation}

\begin{remark} \label{rem1}
In what follows, we assume that~$\omega$ is not an eigenfrequency of the eigenproblem given by~\eqref{eq:strongOri} with $\bff=\bfze$. This ensures the well-posedness of~\eqref{eq:strongOri} and its descendants appearing in the sequel.
\end{remark}

\subsection{Sensitivity Analysis} \label{sub:sensitivity}

\noindent To minimize~\eqref{eq:funcOri}, the idea is to perturb the trial source term $\bff \in C_\delta (\Omega)$ in \eqref{eq:strongOri} by a fixed number $N$ of point sources as
\begin{equation} 
\tilde{\bff}(\bfx) = \bff(\bfx) + \sum_{n=1}^{N} \bfM^{(n)} \sip \nabla_{\!\!\bfx^{(n)}} \delta(\bfx - \bfx^{(n)}), 
\label{eq:fdelta}
\end{equation}
so that $\tilde{\bff} \in C_\delta(\Omega)$ is the perturbed source term. From \eqref{eq:strongZ} and  \eqref{eq:fdelta}, we can introduce the perturbed forward solution $\tilde{\bfu}\!: \Omega \mapsto \mathds{C}^{3}$ as 
\begin{equation}
\left\{
\begin{array}{rllcl}
-\nabla\sip \bfsig[\tilde{\bfu}] - \rho \exs \omega^2 \tilde{\bfu} & = & \tilde{\bff} & \text{in} & \Omega, \\
\tilde{\bfu} & = & \bfze & \mbox{on} & \GamD , \\
\bfn \sip \bfsig[\tilde{\bfu}] & = & \bfze & \text{on} & \GamN, 
\end{array} \right.
\label{eq:strongPer}
\end{equation}
which yields the affiliated misfit functional as 
\begin{equation}
\mathcal{J}(\tilde{\bfu}) = \tfrac{1}{2} \int_{\Gamobs} \! \mathfrak{s}(\bfx) \, \|\tilde{\bfu}-\bfuobs\|^2  \dd\Gamma.
\label{eq:funcPer}
\end{equation}

\paragraph{\bf{Relationship with the topological derivative approach}} In principle, the present treatment could be cast within a broad class of methods based on the concept of topological derivative (TD) \cite{AmmariBook2013,NazarovBook1991,MazyaBook2000} which postulates the nucleation of an infinitesimal new boundary (e.g. a void of a crack), at a prescribed location, in a reference body. The main distinction here is that the majority of TD studies inherently focus on the perturbation of a given cost functional due to a newly created (trial) boundary, when a reference body is subjected to an \emph{external} stimulus. In the context of inverse problems, such configurations are often referred to as \emph{active imaging} configurations, which puts an emphasis on the scattering effect due to a newly created boundary (i.e. scatterer) interacting with a prescribed incident field. By contrast, the present approach links via~\eqref{eq:truef} the TD framework with a source reconstruction problem, where the leading-order perturbation of a cost functional is generated by the strain energy released \emph{during the act of creation} of a new infinitesimal boundary -- as opposed to its interaction with an externally-generated incident field. 

Since~$\bfuobs\!\in\mathds{C}^3$ is interpreted as the Fourier transform of the time traces captured on~$\Gamobs$, the analogous interpretation is applied to the reconstructed point (dipole and double-couple) sources. In this vein, it is convenient to decompose the seismic moment tensor $\bfM^{(n)}\in\mathds{C}^{3\times 3}$ into its (in-phase) real and (out-of-phase) imaginary components as 
\begin{equation}
\bfM^{(n)} = \bfA^{(n)} + i \bfB^{(n)} , \quad \quad \bfA^{(n)},\bfB^{(n)} \in \mathds{R}^{3\times 3}. 
\end{equation}
On writing 
\begin{equation}
\bfA^{(n)} = A^{(n)}_{kl} \bfe_k \otimes \bfe_l, \qquad \bfB^{(n)} = B^{(n)}_{kl} \bfe_k \otimes \bfe_l, \
\end{equation}
the perturbed solution~$\tilde{\bfu}$ can be parsed as
\begin{equation}
\tilde{\bfu}(\bfx) = \bfu(\bfx) + \sum_{n=1}^N \left( A^{(n)}_{kl}\exs \bfv_{kl}^{(n)}(\bfx) + i B^{(n)}_{kl} \exs \bfv_{kl}^{(n)}(\bfx) \right)
\label{eq:solup}
\end{equation}
where $\bfv_{kl}^{(n)}$ ($k,l=\overline{1,3}$) solve the canonical problem
\begin{equation}
\left\{
\begin{array}{rllcl}
-\nabla\sip \bfsig[\bfv_{kl}^{(n)}] - \rho\exs \omega^2 \bfv_{kl}^{(n)} & = & (\bfe_k \!\otimes\! \bfe_l) \sip \nabla_{\!\!\bfx^{(n)}} \delta (\bfx-\bfx^{(n)}) & \text{in} & \Omega , \\
\bfv_{kl}^{(n)} & = & \bfze & \mbox{on} & \GamD , \\
\bfn\sip\bfsig[\bfv_{kl}^{(n)}] & = & \bfze & \text{on} & \GamN.
\end{array}\right.
\label{eq:canonical}
\end{equation}

\begin{remark} 
One may observe that both~$\bfv_{kl}^{(n)}$ and its complex conjugate solve the same boundary value problem thanks to the fact that the source term is real-valued. Since~\eqref{eq:canonical} is well-posed by premise (see Remark~\ref{rem1}), we obtain a key result that $\bfv_{kl}^{(n)}\!: \Omega\mapsto\mathds{R}^3$. In physical terms, this feature reflects the fact that $\bfv_{kl}^{(n)}$ signifies the (in-phase) standing wavefield in a finite elastic body~$\Omega$. For unbounded domains, on the other hand, this conclusion does not hold due to the appearance of the radiation conditions at infinity which (for time-harmonic problems) entail complex-valued coefficients and so phase-varying wavefields.
\end{remark}

\noindent For future reference, we introduce an auxiliary set of vector functions $\bfvv^{(n)}_\ell\!\!:\Omega\mapsto \mathds{R}^3$ ($\ell=\overline{1,6}$) defined as 
\begin{equation}\label{vells}
\begin{aligned}
\bfvv^{(n)}_1 &= \bfv_{11}^{(n)},& \bfvv^{(n)}_2 &=  \bfv_{22}^{(n)},& \bfvv^{(n)}_3 &= \bfv_{33}^{(n)},& \\
\bfvv^{(n)}_4 &= \bfv_{12}^{(n)}+\bfv_{21}^{(n)},& \quad 
\bfvv^{(n)}_5 &= \bfv_{13}^{(n)}+\bfv_{31}^{(n)},& \quad 
\bfvv^{(n)}_6 &= \bfv_{23}^{(n)}+\bfv_{32}^{(n)}.&
\end{aligned}
\end{equation}

On substituting~\eqref{eq:solup} into~\eqref{eq:funcPer}, we obtain
\begin{align}
\mathcal{J}(\tilde{\bfu}) &= \mathcal{J}(\bfu)  \notag \\
&+ \sum_{n=1}^N A_{pq}^{(n)} \int_{\Gamobs} \! \mathfrak{s} \, \bfv_{pq}^{(n)} \sip \Re(\bfu-\bfuobs) \dd\Gamma
 + \tfrac{1}{2} \sum_{m=1}^N \sum_{n=1}^N A_{pq}^{(m)} A_{rs}^{(n)} \int_{\Gamobs} \! \mathfrak{s} \,\bfv_{pq}^{(m)} \sip \bfv_{rs}^{(n)} \dd\Gamma \notag \\ 
&+ \sum_{n=1}^N B_{pq}^{(n)} \int_{\Gamobs} \! \mathfrak{s} \, \bfv_{pq}^{(n)} \sip \Im(\bfu-\bfuobs) \dd\Gamma
 + \tfrac{1}{2} \sum_{m=1}^N \sum_{n=1}^N B_{pq}^{(m)} B_{rs}^{(n)} \int_{\Gamobs} \! \mathfrak{s} \, \bfv_{pq}^{(m)} \sip \bfv_{rs}^{(n)} \dd\Gamma,
\label{eq:expansion}
\end{align}
assuming implicit summation over repeated indexes $p,q,r,s = \overline{1,3}$. 

For a systematic treatment of \eqref{eq:expansion}, we next introduce the vector of \emph{trial} source locations
\begin{equation}
\bfxi = (\bfx^{(1)}, \bfx^{(2)}, \ldots, \bfx^{(N)})^{\mbox{\tiny T}} \in \mathds{R}^{3N},
\end{equation}
and  the affiliated strength vectors
\begin{align}
\bfalpha &= (\bfalpha^{(1)}, \bfalpha^{(2)}, \ldots, \bfalpha^{(N)})^{\mbox{\tiny T}} \in \mathds{R}^{6N},  \\
\bfbeta &= (\bfbeta^{(1)}, \bfbeta^{(2)}, \ldots, \bfbeta^{(N)})^{\mbox{\tiny T}} \in \mathds{R}^{6N},
\end{align}
with  
\begin{align}
\bfalpha^{(n)}&=(A_{11}^{(n)},A_{22}^{(n)},A_{33}^{(n)},A_{12}^{(n)},A_{13}^{(n)},A_{23}^{(n)})^{\mbox{\tiny T}}, \\
\bfbeta^{(n)} &=(B_{11}^{(n)},B_{22}^{(n)},B_{33}^{(n)},B_{12}^{(n)},B_{13}^{(n)},B_{23}^{(n)})^{\mbox{\tiny T}}, 
\end{align}
which take advantage of the symmetry of~$\bfA^{(n)}$ and~$\bfB^{(n)}$.
From these elements, the expansion of \eqref{eq:funcPer} can be rewritten more compactly as
\begin{align}
\Psi(\bfalpha,\bfbeta) 
& = \mathcal{J}(\tilde{\bfu}) - \mathcal{J}(\bfu) \notag \\ 
& = \bfg \sip \bfalpha + \tfrac{1}{2} \bfH \bfalpha \sip \bfalpha + \bfh \sip \bfbeta + \tfrac{1}{2} \bfH \bfbeta \sip \bfbeta.
\label{eq:funcPsi}
\end{align}
With reference to~\eqref{fraks}, vectors $\bfg, \bfh \in \mathds{R}^{6N}$ and matrix $\bfH \in \mathds{R}^{6N\times 6N}$ are given by 
\begin{equation}
\bfg = \left(\begin{array}{c}
\bfg^{(1)} \\ \bfg^{(2)} \\ \vdots \\ \bfg^{(N)}
\end{array}\right), \quad
\bfh = \left(\begin{array}{c}
\bfh^{(1)} \\ \bfh^{(2)} \\ \vdots \\ \bfh^{(N)}
\end{array}\right)\quad \text{and} \quad 
\bfH = \left(\begin{array}{cccc}
\bfH^{(11)} & \bfH^{(12)} & \ldots & \bfH^{(1N)} \\
\bfH^{(21)} & \bfH^{(22)} & \ldots & \bfH^{(2N)} \\
\vdots      & \vdots      & \ddots & \vdots      \\ 
\bfH^{(N1)} & \bfH^{(N2)} & \ldots & \bfH^{(NN)}
\end{array}\right), 
\end{equation}
with the components of $\bfg^{(n)},\bfh^{(n)}\in\mathds{R}^6$ and $\bfH^{(mn)}\!\in\mathds{R}^{6\times 6}$ ($m,n=\overline{1,N}$) defined as 
\begin{align}
g^{(n)}_{\ell} &= \int_{\Gamobs} \!\mathfrak{s}\, \bfvv^{(n)}_{\ell} \sip \exs\Re (\bfu-\bfuobs) \dd\Gamma,  \\
h^{(n)}_{\ell} &= \int_{\Gamobs} \!\mathfrak{s}\, \bfvv^{(n)}_{\ell} \sip \exs\Im (\bfu-\bfuobs) \dd\Gamma, \qquad \ell=\overline{1,6}, 
\end{align}
and  
\begin{align}
H^{(mn)}_{\ell \kappa} = \int_{\Gamobs} \!\mathfrak{s}\, \bfvv^{(m)}_{\ell} \sip \exs \bfvv^{(n)}_{\kappa} \dd\Gamma, \qquad \ell,\kappa=\overline{1,6}. 
\end{align}

\subsection{Reconstruction Algorithm} \label{sub:algorithm}

\noindent The micro-seismic fault reconstruction algorithm has been introduced in \cite{AmadIJSS2020} within the context of two-dimensional elastodynamic problems. In this work, these ideas are extended to the inverse (seismic) source reconstruction problem into three spatial dimensions. 

Note that $\Psi$ from \eqref{eq:funcPsi} is a quadratic form with respect to $\bfalpha$ and $\bfbeta$. After applying the optimality conditions with respect to $(\bfalpha,\bfbeta)$, for each fixed par $(N,\bfxi)$, we retrieve the following linear systems
\begin{equation}
\bfH \bfalpha = -\bfg \quad \text{and} \quad \bfH \bfbeta = -\bfh.
\label{eq:systems}
\end{equation}
The solutions to the above linear systems become implicit functions of the locations $\bfxi$, namely $\bfalpha = \bfalpha(\bfxi)$ and $\bfbeta = \bfbeta(\bfxi)$. Therefore, substituting \eqref{eq:systems} in \eqref{eq:funcPsi}, we can define a minimization problem with respect to the locations $\bfxi$, namely
\begin{equation}
\bfxi^\star = \underset{\bfxi \subset X}{\mathrm{argmin}} \left\{\Psi(\bfalpha(\bfxi),\bfbeta(\bfxi)) = 
\tfrac{1}{2}(\bfg \sip \bfalpha(\bfxi) + \bfh \sip \bfbeta(\bfxi))\right\}.
\label{eq:optimal}
\end{equation}
The optimal vector of source locations $\bfxi^\star$ can be obtained via a combinatorial search over a prescribed grid $X$ of trial source locations. For more specialized search procedures from the computational point of view, see e.g. \cite{MachadoM2AS2017,AmadIJSS2020}.

Now, we have all elements to introduce a second-order reconstruction algorithm. With reference to Algorithm~\ref{algoritmo1} listed below, the input to the inverse analysis are the number $N$ of micro-seismic point sources, the grid $X$ of trial source locations, and the canonical solutions $\bfv_{kl}^{(n)}$ evaluated at each grid point $\bfx^{(n)} \!\in\! X$, $n\!=\!\overline{1,N}$. The algorithm returns the optimal source locations $\bfxi^\star$, the respective moment tensor components in terms of optimal vectors $\bfalpha^\star = \bfalpha(\bfxi^\star)$ and $\bfbeta^\star = \bfbeta(\bfxi^\star)$ and the objective functional value $\Psi^\star = \Psi(\bfalpha^\star, \bfbeta^\star)$. The associated reconstructed $n$th moment tensor is denoted as $\bfM^{(n)}\rec$.
\begin{algorithm}\caption{Micro-seismic fault reconstruction.}
\begin{algorithmic}[1]
\SetAlgoLined
\STATE \textbf{input :} $N$, $X$, $\bfH$, $\bfg$, $\bfh$;
\STATE \textbf{output:} the optimal solution $\bfxi^\star$, $(\bfalpha^\star,\bfbeta^\star)$, $\Psi^\star$;
\STATE \textbf{initialization:} $\bfxi^\star \leftarrow \bfze$, $(\bfalpha^\star,\bfbeta^\star)  \leftarrow (\bfze,\bfze)$, $\Psi^\star \leftarrow \infty$;
\STATE \textbf{for} $n_1 \leftarrow 1$ to $\mathrm{card}\{X\}$ do
\STATE \hspace{0.5cm} \textbf{for} $n_2 \leftarrow n_{1}+1$ to $\mathrm{card}\{X\}$ do
\STATE \hspace{1.0cm} \vdots 
\STATE \hspace{1.0cm} \textbf{for} $n_N \leftarrow n_{N-1}+1$ to $\mathrm{card}\{X\}$ do
\STATE \begin{equation*}
\hspace{1.8cm} \bfg \leftarrow \left(\begin{array}{c}
\bfg^{(n_1)}\\
\bfg^{(n_2)}\\
\vdots      \\
\bfg^{(n_N)}
\end{array}\right), \;
\bfh \leftarrow \left(\begin{array}{c}
\bfh^{(n_1)}\\
\bfh^{(n_2)}\\
\vdots      \\
\bfh^{(n_N)}
\end{array}\right), \;
\bfH \leftarrow
\left(\begin{array}{cccc}
\bfH^{(n_1,n_1)} & \bfH^{(n_1,n_2)} & \cdots & \bfH^{(n_1,n_N)} \\
\bfH^{(n_2,n_1)} & \bfH^{(n_2,n_2)} & \cdots & \bfH^{(n_2,n_N)} \\
\vdots           & \vdots           & \ddots & \vdots           \\
\bfH^{(n_N,n_1)} & \bfH^{(n_N,n_2)} & \cdots & \bfH^{(n_N,n_N)}
\end{array}\right);
\end{equation*}
\STATE \hspace{1.8cm} $\bfalpha \leftarrow -\bfH^{-1}\bfg$, $\bfbeta \leftarrow -\bfH^{-1}\bfh$, 
$\Psi \leftarrow \tfrac{1}{2}(\bfg \sip \bfalpha + \bfh \sip \bfbeta)$;
\STATE \hspace{1.8cm} $\mathcal{I} \leftarrow (n_1,n_2, \cdots, n_N)$, $\bfxi \leftarrow \Pi(\mathcal{I})$;
\STATE \hspace{1.8cm} \textbf{if} $\Psi < \Psi^\star$ then
\STATE \hspace{2.3cm} $\bfxi^\star \leftarrow \bfxi$, $(\bfalpha^\star,\bfbeta^\star) \leftarrow (\bfalpha,\bfbeta)$, $\Psi^\star \leftarrow \Psi$;
\STATE \hspace{1.8cm} \textbf{end if}
\STATE \hspace{1.0cm} \textbf{end for}
\STATE \hspace{0.5cm} \textbf{end for} 
\STATE \textbf{end for} 
\end{algorithmic}
\label{algoritmo1}
\end{algorithm}

\section{Numerical Results} \label{sec:numerics}

\begin{figure}[H]
	\center
	\subfigure[]{\includegraphics[height=3.9cm]{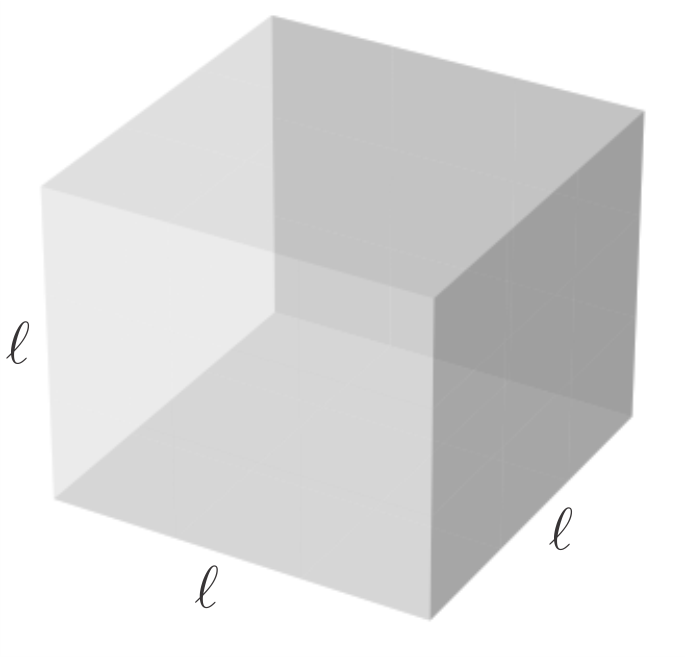}\label{fig:block}}\qquad
	\subfigure[]{\includegraphics[height=3.9cm]{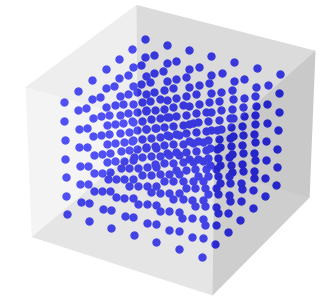}\label{fig:source}}
	\caption{Working domain: (a) cube-shaped elastic body and (b) set of admissible point source locations.}
\end{figure}

\begin{figure}[H]
	\center
	\subfigure[]{\includegraphics[height=3.9cm]{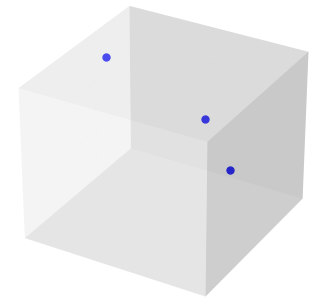}\label{fig:sensor3}} \qquad
        \subfigure[]{\includegraphics[height=3.9cm]{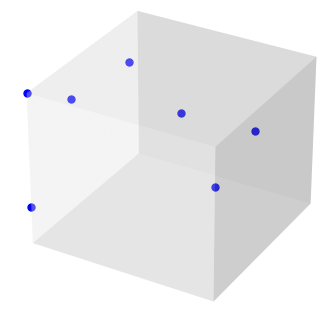}\label{fig:sensor7}} \qquad
        \subfigure[]{\includegraphics[height=3.9cm]{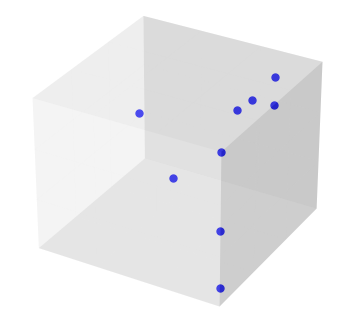}\label{fig:sensor9}}
	\caption{Surface sensing configuration featuring: (a) three accelerometers, (b) seven accelerometers and (c) nine accelerometers.}
\label{fig:pbo}
\end{figure}

\noindent The elastic body $\Omega$ used for numerical simulations is taken as a cube-shaped block of dimensions $\ell\times\ell\times\ell$, as shown in Fig.~\ref{fig:block}, which is fixed on the bottom face. The pointwise tri-axial motion sensors are assumed to be attached along the boundary $\Gamobs$ as described in the sequel (see Fig.~\ref{fig:pbo}). The dimensionless frequency of acoustic emission is taken as 
\begin{equation}
\frac{\omega \ell}{\sqrt{\mu/\rho}} = 10 \pi,
\label{eq:frequency}
\end{equation}
resulting in the specimen-size-to-shear-wavelength ratio of $\ell/ \lambda_s=5$. 
%

The forward elastodynamic problem is solved via Galerkin finite element method by using the open software package \textit{Netgen/NGSolve} \cite{SchoberlCVS1997}. The cube-shaped domain from Fig.~\ref{fig:block} is split into 512 equisized cubes. The set of admissible locations $X$ is obtained by selecting $343$ interior vertices from the resulting grid, see Fig.~\ref{fig:source}. In order to fulfill the numerical condition presented in \cite{BabuskaCMA1995}, each small cube is divided into tetrahedral finite elements, leading to $196608$ elements and $35937$ nodes. For the purpose of numerical simulations, the dipoles are approximated by a Gaussian distribution
\begin{equation}
\delta(\bfx-\bfx^{(n)}) = \underset{\varepsilon \rightarrow 0}{\lim} \;
\frac{1}{\varepsilon^3} \exp{\left(-\frac{\| \bfx - \bfx^{(n)} \|^2}{2 \varepsilon^2}\right)},
\end{equation}
whose gradient with respect to $\bfx^{(n)}$ can be obtained as
\begin{equation}
\nabla_{\bfx^{(i)}}\delta(\bfx-\bfx^{(n)}) = \underset{\varepsilon \rightarrow 0}{\lim} \;
\frac{\bfx-\bfx^{(n)}}{\varepsilon^5} \exp{\left(-\frac{\| \bfx - \bfx^{(n)} \|^2}{2 \varepsilon^2}\right)}.
\end{equation}
In the sequel, we assume $\varepsilon = 10^{-2} \times h^e$, where $h^e$ denotes the size of the smallest finite element. 

Finally, a number of $3$, $7$, and $9$ triaxial accelerometers are distributed along the boundary of the elastic body as shown in Fig.~\ref{fig:sensor3}, \ref{fig:sensor7}, and \ref{fig:sensor9}, respectively. After performing an exhaustive numerical study in the idealized scenario with no noise and assuming $\widehat{\bfxi} \subset X$, we note that at least $2N+1$ sensors are needed to reconstruct perfectly the locations and associated moment tensors of a number $N$ of faults. 

\subsection{Reconstruction of a single event.}  

\noindent In the first example, we deal with a selection of moment tensors and affiliated focal mechanisms described in the book by \citeauthor{SteinBook2003} \citep{SteinBook2003}. We consider an isotropic elastic block with $\ell=0.3$, Poisson's ratio $\nu=0.3$ and triaxial accelerometers attached to its surface according to Fig.~\ref{fig:sensor3}. The micro-seismic event location to be reconstructed is given by $\widehat{\bfx} = (0.15, 0.15,  0.15)$. The featured moment tensors (normalized to unit magnitude) and their fault mechanisms (pictorially described by the beach-ball scheme) are listed in Table \ref{tab:noise1}.
\begin{table}[H]
\center
\caption{Identification of a single micro-seismic event at $\widehat{\bfx} = (0.15, 0.15,  0.15)$, where $\widehat{\bfM}$ and $\bfM\rec$ denote respectively the ``true'' and reconstructed moment tensors. 
The system north, west, and up (NWU) is used to represent the moment tensors as in \cite{SteinBook2003}.}
\medskip
\begin{longtabu}{r c c | c c c}
\tabucline[1pt]{-} \addlinespace[2pt]
Moment Tensor & $\widehat{\bfM}$ & $\bfM\rec$ & 
Moment Tensor & $\widehat{\bfM}$ & $\bfM\rec$ \\ \addlinespace[2pt] \tabucline[1pt]{-} 
$\frac{1}{\sqrt{3}}\left(\begin{array}{rrr}
1&0&0\\
0&1&0\\
0&0&1
\end{array}\right)$  & 
\parbox[c]{4em}{\includegraphics[scale=0.25]{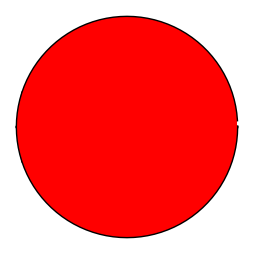}}  & 
\parbox[c]{4em}{\includegraphics[scale=0.25]{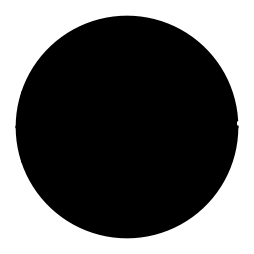}} & 
$-\frac{1}{\sqrt{3}}\left(\begin{array}{rrr}
1&0&0\\
0&1&0\\
0&0&1
\end{array}\right)$  & 
\parbox[c]{4em}{\includegraphics[scale=0.25]{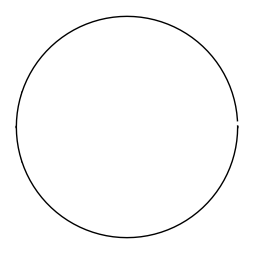}}  & 
\parbox[c]{4em}{\includegraphics[scale=0.25]{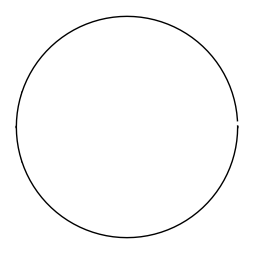}} \\ 
$-\frac{1}{\sqrt{2}}\left(\begin{array}{rrr}
0&1&0\\
1&0&0\\
0&0&0
\end{array}\right)$  &
\parbox[c]{4em}{\includegraphics[scale=0.25]{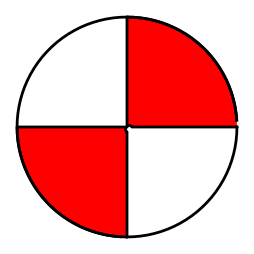}}  & 
\parbox[c]{4em}{\includegraphics[scale=0.25]{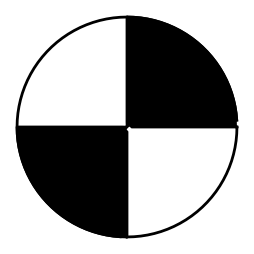}} 
& 
$\frac{1}{\sqrt{2}}\left(\begin{array}{rrr}
1&0&0\\
0&-1&0\\
0&0&0
\end{array}\right)$  &
\parbox[c]{4em}{\includegraphics[scale=0.25]{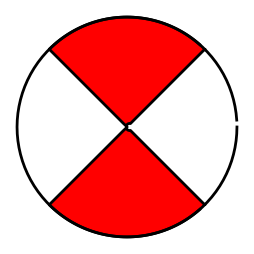}}  & 
\parbox[c]{4em}{\includegraphics[scale=0.25]{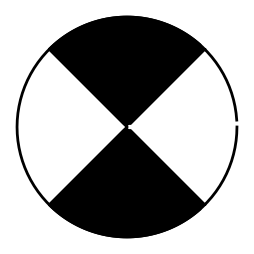}} 
\\ 
$-\frac{1}{\sqrt{2}}\left(\begin{array}{rrr}
0&0&1\\
0&0&0\\
1&0&0
\end{array}\right)$  &
\parbox[c]{4em}{\includegraphics[scale=0.25]{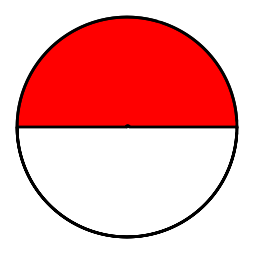}}  & 
\parbox[c]{4em}{\includegraphics[scale=0.25]{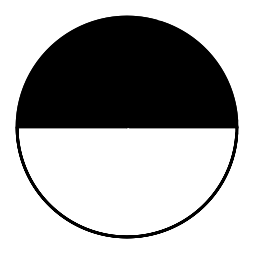}} 
& 
$-\frac{1}{\sqrt{2}}\left(\begin{array}{rrr}
0&0&0\\
0&0&1\\
0&1&0
\end{array}\right)$  &
\parbox[c]{4em}{\includegraphics[scale=0.25]{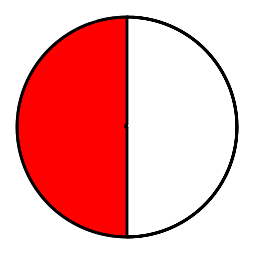}}  & 
\parbox[c]{4em}{\includegraphics[scale=0.25]{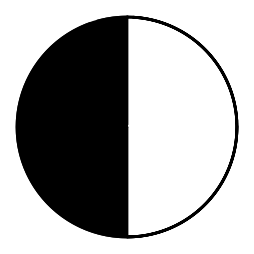}} 
\\ 
$\frac{1}{\sqrt{2}}\left(\begin{array}{rrr}
-1&0&0\\
0&0&0\\
0&0&1
\end{array}\right)$  &
\parbox[c]{4em}{\includegraphics[scale=0.25]{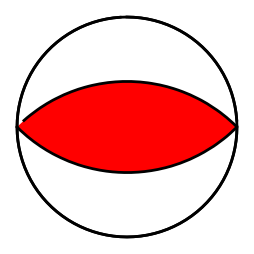}}  & 
\parbox[c]{4em}{\includegraphics[scale=0.25]{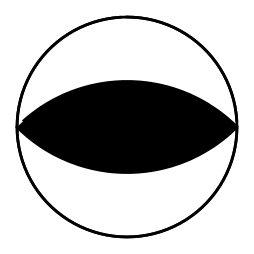}} 
& 
$\frac{1}{\sqrt{2}}\left(\begin{array}{rrr}
0&0&0\\
0&-1&0\\
0&0&1
\end{array}\right)$  &
\parbox[c]{4em}{\includegraphics[scale=0.25]{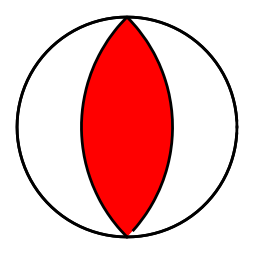}}  & 
\parbox[c]{4em}{\includegraphics[scale=0.25]{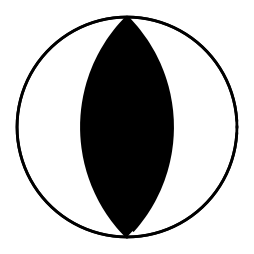}} 
\\ 
$\frac{1}{\sqrt{6}}\left(\begin{array}{rrr}
1&0&0\\
0&-2&0\\
0&0&1
\end{array}\right)$  &
\parbox[c]{4em}{\includegraphics[scale=0.25]{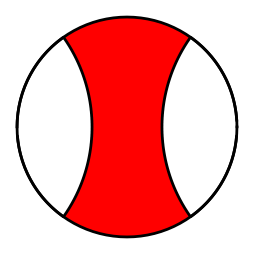}}  & 
\parbox[c]{4em}{\includegraphics[scale=0.25]{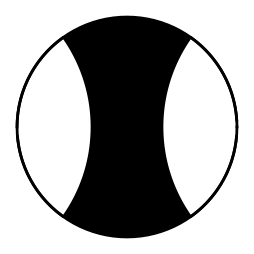}} 
& 
$\frac{1}{\sqrt{6}}\left(\begin{array}{rrr}
-2&0&0\\
0&1&0\\
0&0&1
\end{array}\right)$  &
\parbox[c]{4em}{\includegraphics[scale=0.25]{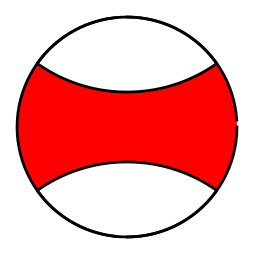}}  & 
\parbox[c]{4em}{\includegraphics[scale=0.25]{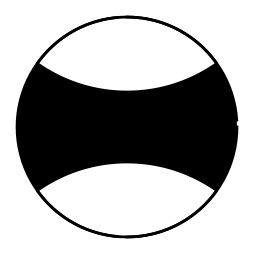}}
\\ 
$\frac{1}{\sqrt{6}}\left(\begin{array}{rrr}
1&0&0\\
0&1&0\\
0&0&-2
\end{array}\right)$  &
\parbox[c]{4em}{\includegraphics[scale=0.25]{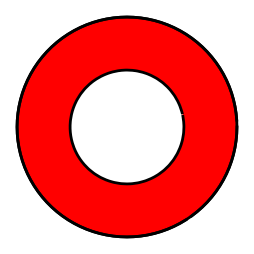}}  & 
\parbox[c]{4em}{\includegraphics[scale=0.25]{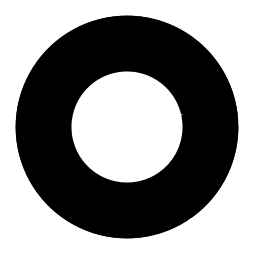}}
& 
$-\frac{1}{\sqrt{6}}\left(\begin{array}{rrr}
1&0&0\\
0&1&0\\
0&0&-2
\end{array}\right)$  &
\parbox[c]{4em}{\includegraphics[scale=0.25]{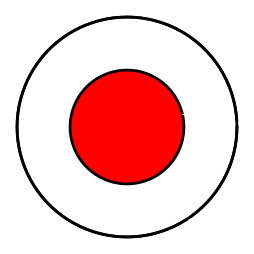}}  & 
\parbox[c]{4em}{\includegraphics[scale=0.25]{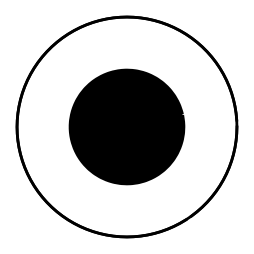}}  \\ 
\tabucline[1pt]{-}
\end{longtabu}
\label{tab:noise1}
\end{table}

\subsection{Reconstruction of three contemporaneous events.}

\begin{figure}[H]
\begin{center}
\includegraphics[scale = 1.0]{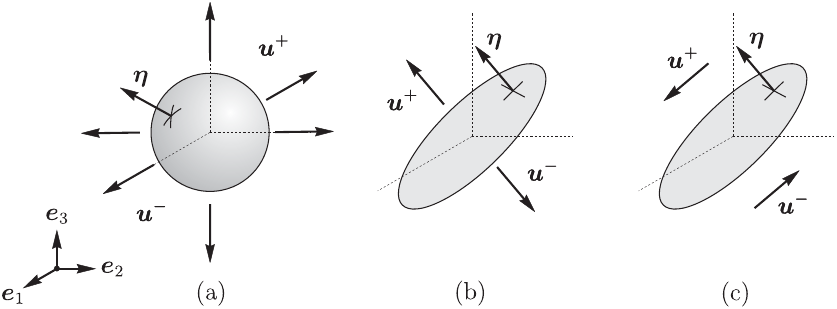} 
\end{center}
\caption{Elemental types of micro-seismic events: (a) cavitation, (b) mode I crack, and (c) and mode II crack.}
\label{fig:seismic3d}
\end{figure}

\noindent In the next example, we consider the reconstruction of cavitation, tensile, and shear events overlapping in time. A description of the featured micro-seismic events is shown in Fig.~\ref{fig:seismic3d}. A generic seismic moment tensor $\bfM$ can be written as
\begin{equation}
\bfM = a_\Xi \bfC([\![\bfu]\!] \odot \bfeta) = a_\Xi (2\mu ([\![\bfu]\!] \odot \bfeta) + \lambda ([\![\bfu]\!] \sip \bfeta) \bfI),
\end{equation}
where $a_\Xi$ is the area of a newly created micro-fault $\Xi$ giving rise to the acoustic emission; $\bfeta$ is the unit normal to the micro-fault surface; $[\![\bfu]\!]$ is the average displacement jump across the micro-fault; $\bfC$ the elasticity tensor, and $\bfu \odot \bfv$ is the symmetrized tensor product between vectors $\bfu$ and $\bfv$, namely
\begin{equation}
\bfu \odot \bfv = \tfrac{1}{2}(\bfu \otimes \bfv + \bfv \otimes \bfu).
\end{equation}
The quantity $a_\Xi$ is replaced by the \emph{fault strength} $\gamma \in \mathds{C}$ to account for out-of-phase events. Three micro-seismic moment tensors are considered according to the sketch in Fig.~\ref{fig:seismic3d} and one additional which combines two of them, namely: 
\begin{enumerate}
\item [(a)] \textbf{Cavitation:} For $\bfeta = [\![\bfu]\!] = \bfe_i$, with $i \in \overline{1,3}$, there is
\begin{equation}
\widehat{\bfM} = \gamma\begin{pmatrix}
2 \mu + 3 \lambda & 0 & 0 \\
0 & 2 \mu + 3 \lambda & 0 \\
0 & 0 & 2 \mu + 3 \lambda
\end{pmatrix};
\label{eq:cavitation}
\end{equation}
\item [(b)] \textbf{Tensile crack:} For $\bfeta = (1,0,0)$ and $[\![\bfu]\!] = (1,0,0)$, we have  
\begin{equation}
\widehat{\bfM} =\gamma\begin{pmatrix}
2 \mu + \lambda & 0 & 0 \\
0 & \lambda & 0 \\
0 & 0 & \lambda
\end{pmatrix};
\label{eq:traction}
\end{equation}
\item [(c)] \textbf{Shear crack:}  For $\bfeta = (1,0,0)$ and $[\![\bfu]\!] = (0,1,0)$, we have  
\begin{equation}
\widehat{\bfM} = \gamma\begin{pmatrix}
0 & 2 \mu  & 0 \\
2 \mu  & 0 & 0 \\
0 & 0 & 0
\end{pmatrix};
\label{eq:shear}
\end{equation}
\item [(d)] \textbf{Mixed-mode crack:}  By combining \eqref{eq:traction} and \eqref{eq:shear}, we have
\begin{equation}
\widehat{\bfM} = \gamma\begin{pmatrix}
2\mu+\lambda & 2 \mu  & 0 \\
2 \mu  & \lambda & 0 \\
0 & 0 & \lambda
\end{pmatrix}.
\label{eq:shear-trac}
\end{equation}
\end{enumerate}
In this example, we consider an elastic block with the Poisson's ratio $\nu=0.2$. In physical terms, this configuration could correspond to e.g.~a block of sandstone with $\ell\!=\!0.16$m, $E\!=\!20$GPa, $\rho\!=\!2300$kg/m$^3$ and, according to~\eqref{eq:frequency}, the frequency of acoustic emission $\!f\!=\omega/(2\pi) \!\approx\! 60$kHz \cite{shengxiang2021study}.
 
Various external factors may generate meaningful changes in the dynamic response of the point-wise sensors used to measure $\bfuobs$ on $\Gamobs$, such as acoustic noise, temperature variations, and humidity. In order to verify the robustness of the method with respect to noisy data, the measurements $\bfuobs$ are corrupted with white Gaussian noise (WGN). In particular, the corrupted measurement $\bfuobs_\zeta$ is given by
\begin{equation}
\bfuobs_\zeta(x) = \bfuobs(x)(1 + \zeta \phi(x)),
\end{equation}
where $\phi$ is a random variable with a uniform distribution over $[0, 1)$, and $\zeta$ is the noise level. Another important feature of an AE event exposed by the reconstructed moment tensor $\boldsymbol{M}^{(n)}\rec$ is its magnitude, reflected in the Frobenius norm $\|\boldsymbol{M}^{(n)}\rec\|$. The latter quantity provides an implicit information about the ``importance" of an event, i.e. the size of the newly created surface area $a_{\Xi}^{(n)}$. In this vein, we introduce a relative error function of the form
\begin{equation}
\mathcal{E}^{(n)}\rec = \frac{\|\widehat{\bfM}^{(n)}-\bfM\rec^{(n)}\|}{\|\widehat{\bfM}^{(n)}\|}.
\label{eq:error-rec}
\end{equation}

\subsubsection{Contemporaneous on-grid events $\widehat{\bfx}^{(n)} \in X$.}

\noindent We aim to reconstruct three micro-seismic sources representing cavitation, tensile crack, and shear crack events. The locations $\widehat{\bfx}^{(n)}$ and intensities $\gamma^{(n)}$ are reported in Table \ref{tab:example2}. 

\begin{figure}[H]   
    \center
    \subfigure[$0\%$]{\includegraphics[scale=0.5]{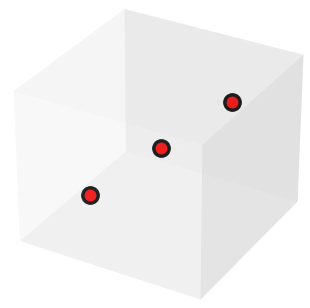}\label{fig:ex2Ar0}}\qquad
    \subfigure[$3\%$]{\includegraphics[scale=0.5]{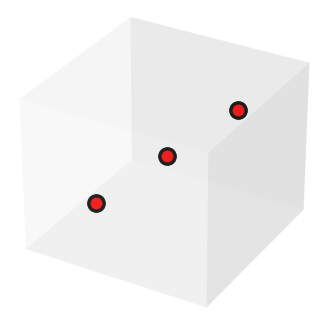}\label{fig:ex2Ar3}}\qquad
    \subfigure[$4\%$]{\includegraphics[scale=0.5]{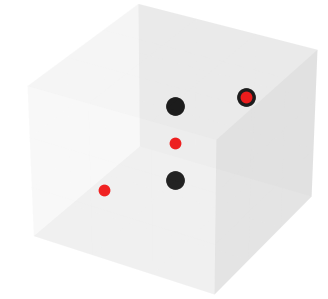}\label{fig:ex2Ar4}}
    \caption{Contemporaneous \emph{on-grid} events: Multi-event location results for $0\%$, $3\%$ and $4\%$ of noise, assuming 7 sensors distributed as in Fig.~\ref{fig:sensor7}.}
    \label{fig:example2locationA}
\end{figure}

\begin{table}[H]
    \centering
    \caption{Contemporaneous on-grid micro-seismic sources.}
    \begin{tabu}{cccc} \tabucline[1pt]{-} \addlinespace[2pt]
         $n$th  & event &  $\gamma^{(n)}$ & $\widehat{\bfx}^{(n)}$ \\ 
         \addlinespace[2pt] \tabucline[1pt]{-} \addlinespace[2pt]
         $1$ & shear      &  $(1.0+2.0i)\times 10^{-10}$ & $(0.12, 0.12, 0.12)$ \\  
         $2$ & cavitation &  $(1.0+1.0i)\times 10^{-10}$ & $(0.04, 0.04, 0.04)$ \\ 
         $3$ & tensile    &  $(2.0+1.0i)\times 10^{-10}$ & $(0.08, 0.08, 0.08)$ \\
         \tabucline[1pt]{-}
    \end{tabu}
    \label{tab:example2}
\end{table}
We start by considering $7$ sensors as reported in Fig.~\ref{fig:sensor7}. The reconstructed fault locations for $0\%$, $3\%$ and $4\%$ of noise are shown in Fig.~\ref{fig:example2locationA}. The true (target) and reconstructed micro-faults locations are represented graphically by (smaller) red and (larger) black balls, respectively. The found locations are exact up to $3\%$ of noise. However, the locations of two faults are missed for $4\%$ of noise, as shown in Fig.~\ref{fig:ex2Ar4}.

\begin{figure}[H]   
    \center
    \subfigure[$4\%$]{\includegraphics[scale=0.5]{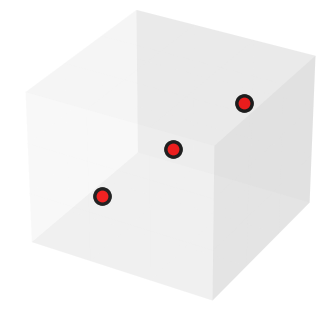}\label{fig:ex2r4}}\qquad
    \subfigure[$8\%$]{\includegraphics[scale=0.5]{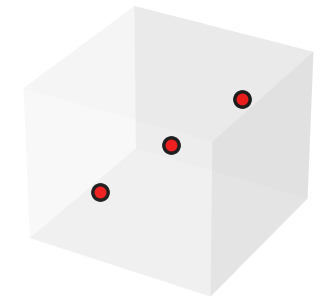}\label{fig:ex2r8}}\qquad
    \subfigure[$9\%$]{\includegraphics[scale=0.5]{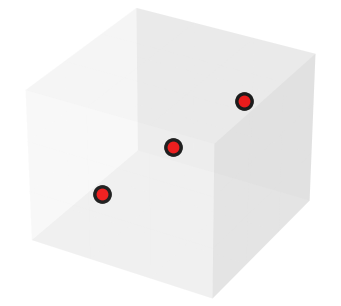}\label{fig:ex2r9}}\qquad
    \subfigure[$10\%$]{\includegraphics[scale=0.5]{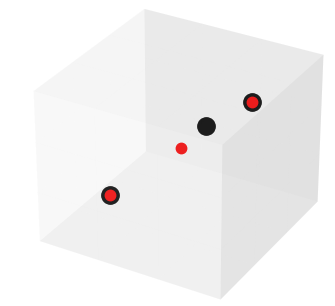}\label{fig:ex2r10}}
    \caption{Contemporaneous \emph{on-grid} events: Multi-event location results using $4\%$, $8\%$, $9\%$ and $10\%$ of noise, assuming 9 sensors distributed as in Fig.~\ref{fig:sensor9}.}
    \label{fig:example2location}
\end{figure}

Next, we consider $9$ sensors as shown in Fig.~\ref{fig:sensor9}. The reconstructed fault locations for $4\%$, $8\%$, $9\%$ and $10\%$ of noise are shown in Fig.~\ref{fig:example2location}. The true (target) and reconstructed micro-faults locations are represented graphically by (smaller) red and (larger) black balls, respectively. The fault locations in Fig.~\ref{fig:example2location} are visually exact up to $9\%$ of noise. However, event location for one out of the three micro-faults shows an error at $10\%$ of noise. The results indicate that the proposed reconstruction scheme is (for the present sensing configuration) very robust. 
The focal mechanisms featured by the reconstructed moment tensors representing the cavitation, tensile crack, and shear crack events are shown in Table \ref{tab:ex2cavsheatens} and \ref{tab:ex2cavsheatensimag} in terms of the (in-phase) real and (out-of-phase) imaginary components, respectively. As can be seen from the tables, the reconstructed focal mechanisms for both tensile and shear crack events retain reasonable veracity for the noise levels of up $2\%$. By contrast, reconstruction of the cavitation (i.e. isotropic dilation) event seems to be very sensitive to measurement noise and effectively fails at $\zeta=2\%$ and beyond. 
\begin{table}[H]
\center
\caption{Contemporaneous \emph{on-grid} events: Real component of the reconstructed source mechanisms for the cavitation, tensile crack, and shear crack event versus the noise level.}
\begin{tabu}{cccccccc} \tabucline[1pt]{-}
target  & $\zeta = 0\%$ & $\zeta = 2\%$ & $\zeta = 4\%$  & $\zeta = 6\%$& $\zeta = 8\%$ & $\zeta = 9\%$ & $\zeta = 10\%$ \\ 
\tabucline[1pt]{-}
\parbox[c]{4em}{\includegraphics[scale=0.25]{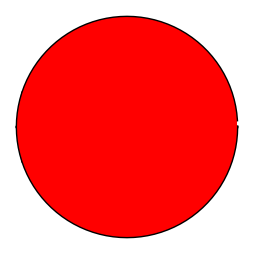}}  & 
\parbox[c]{4em}{\includegraphics[scale=0.25]{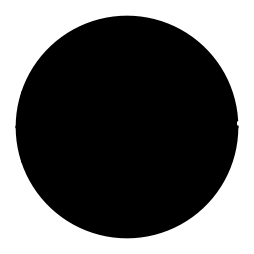}} & 
\parbox[c]{4em}{\includegraphics[scale=0.25]{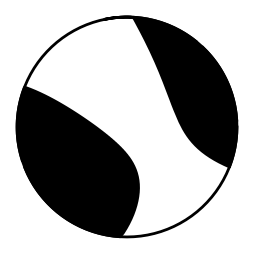}} &
\parbox[c]{4em}{\includegraphics[scale=0.25]{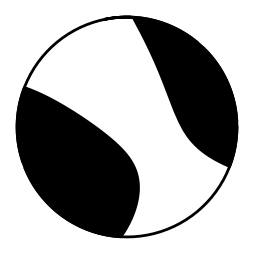}} & 
\parbox[c]{4em}{\includegraphics[scale=0.25]{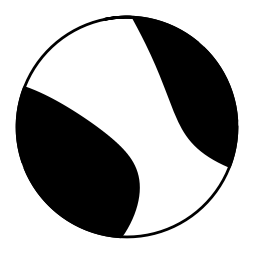}} &
\parbox[c]{4em}{\includegraphics[scale=0.25]{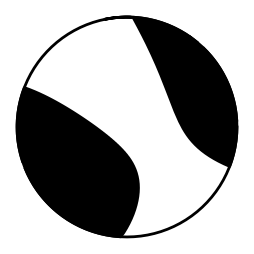}} &
\parbox[c]{4em}{\includegraphics[scale=0.25]{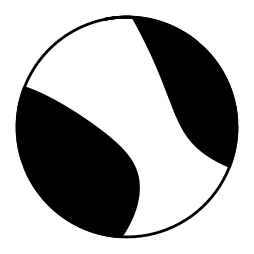}} &
\parbox[c]{4em}{\includegraphics[scale=0.25]{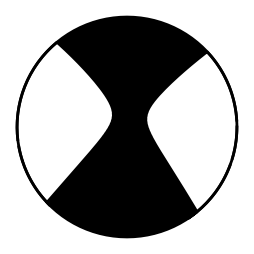}} \\
\parbox[c]{4em}{\includegraphics[scale=0.25]{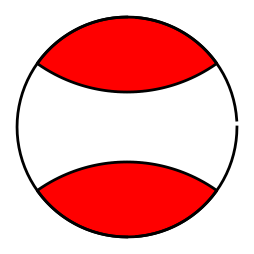}}  & 
\parbox[c]{4em}{\includegraphics[scale=0.25]{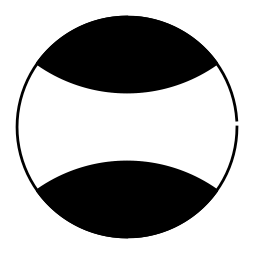}} & 
\parbox[c]{4em}{\includegraphics[scale=0.25]{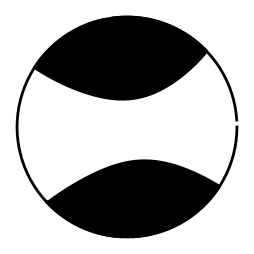}} &
\parbox[c]{4em}{\includegraphics[scale=0.25]{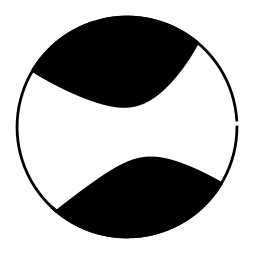}} & 
\parbox[c]{4em}{\includegraphics[scale=0.25]{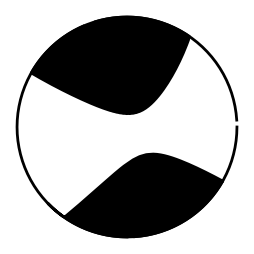}} &
\parbox[c]{4em}{\includegraphics[scale=0.25]{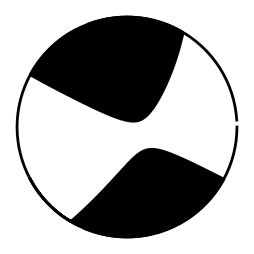}} &
\parbox[c]{4em}{\includegraphics[scale=0.25]{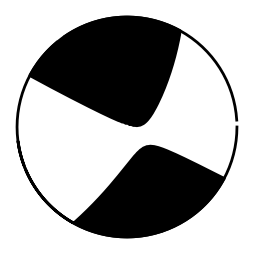}} &
\parbox[c]{4em}{\includegraphics[scale=0.25]{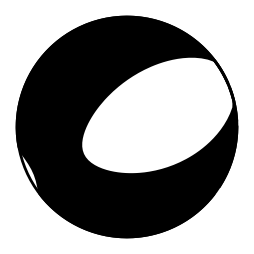}}\\
\parbox[c]{4em}{\includegraphics[scale=0.25]{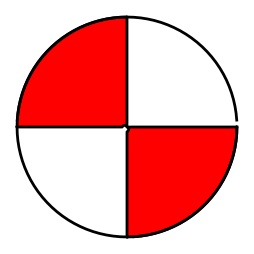}}  &  
\parbox[c]{4em}{\includegraphics[scale=0.25]{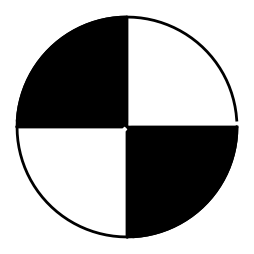}} & 
\parbox[c]{4em}{\includegraphics[scale=0.25]{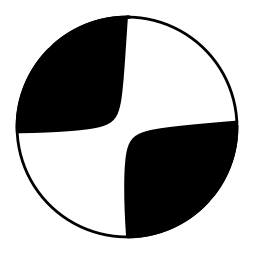}} &
\parbox[c]{4em}{\includegraphics[scale=0.25]{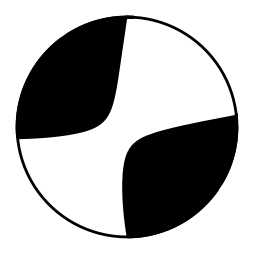}} & 
\parbox[c]{4em}{\includegraphics[scale=0.25]{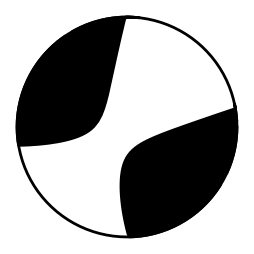}} &
\parbox[c]{4em}{\includegraphics[scale=0.25]{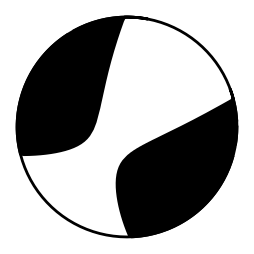}} &
\parbox[c]{4em}{\includegraphics[scale=0.25]{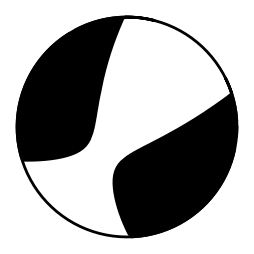}} &
\parbox[c]{4em}{\includegraphics[scale=0.25]{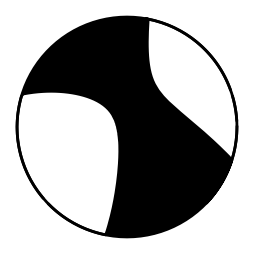}} \\
\tabucline[1pt]{-}
\end{tabu}
\label{tab:ex2cavsheatens}
\end{table}
\begin{table}[H]
\center
\caption{Contemporaneous \emph{on-grid} events: Imaginary component of the reconstructed source mechanisms for the cavitation, tensile crack, and shear crack event versus the noise level.}
\begin{tabu}{cccccccc} \tabucline[1pt]{-}
target  & $\zeta = 0\%$ & $\zeta = 2\%$ & $\zeta = 4\%$  & $\zeta = 6\%$& $\zeta = 8\%$ & $\zeta = 9\%$ & $\zeta = 10\%$ \\ 
\tabucline[1pt]{-}
\parbox[c]{4em}{\includegraphics[scale=0.25]{figures/example2-ongrid/cav1ruido0t.png}}  &  
\parbox[c]{4em}{\includegraphics[scale=0.25]{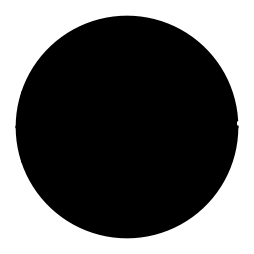}} & 
\parbox[c]{4em}{\includegraphics[scale=0.25]{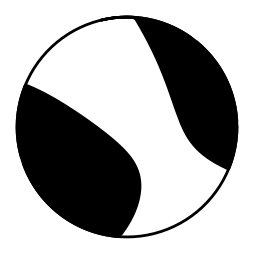}} &
\parbox[c]{4em}{\includegraphics[scale=0.25]{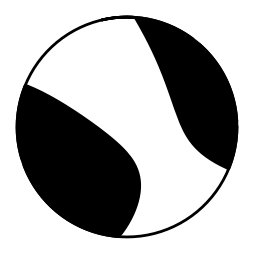}} & 
\parbox[c]{4em}{\includegraphics[scale=0.25]{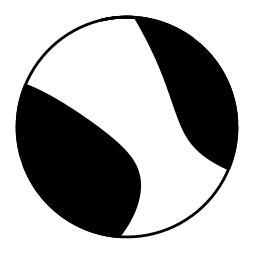}} &
\parbox[c]{4em}{\includegraphics[scale=0.25]{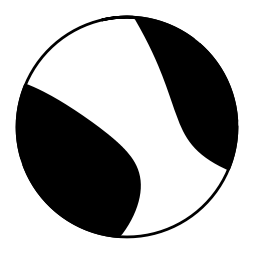}} &
\parbox[c]{4em}{\includegraphics[scale=0.25]{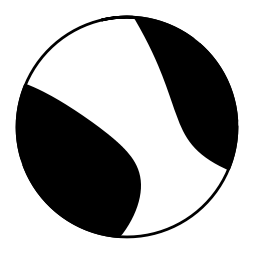}} &
\parbox[c]{4em}{\includegraphics[scale=0.25]{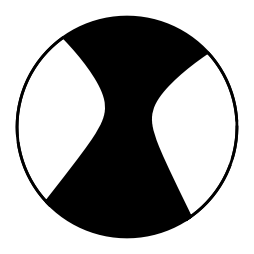}} \\
\parbox[c]{4em}{\includegraphics[scale=0.25]{figures/example2-ongrid/trac1ruido1t.png}}   & 
\parbox[c]{4em}{\includegraphics[scale=0.25]{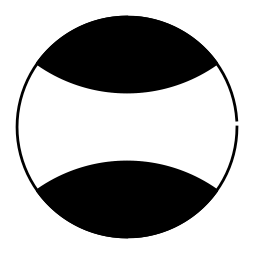}} & 
\parbox[c]{4em}{\includegraphics[scale=0.25]{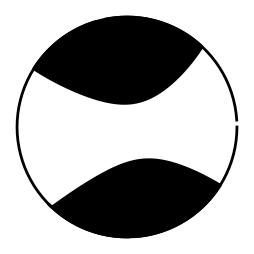}} &
\parbox[c]{4em}{\includegraphics[scale=0.25]{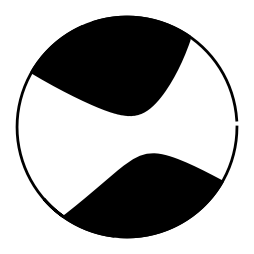}} & 
\parbox[c]{4em}{\includegraphics[scale=0.25]{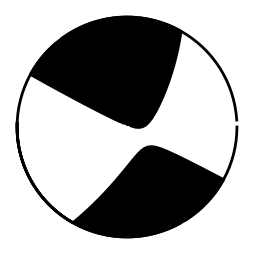}} &
\parbox[c]{4em}{\includegraphics[scale=0.25]{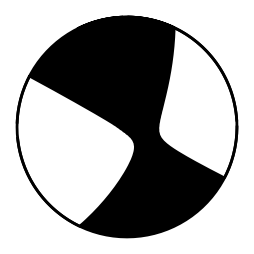}} &
\parbox[c]{4em}{\includegraphics[scale=0.25]{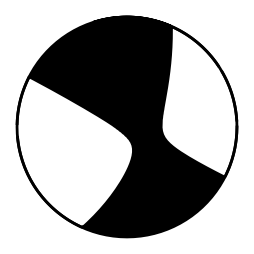}} &
\parbox[c]{4em}{\includegraphics[scale=0.25]{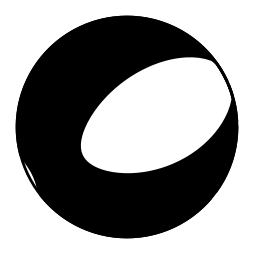}} \\
\parbox[c]{4em}{\includegraphics[scale=0.25]{figures/example2-ongrid/shear1ruido0t.png}}   & 
\parbox[c]{4em}{\includegraphics[scale=0.25]{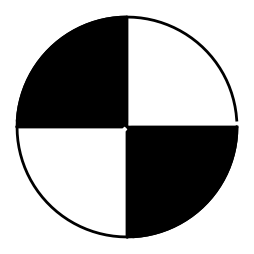}} & 
\parbox[c]{4em}{\includegraphics[scale=0.25]{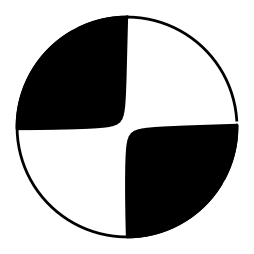}} &
\parbox[c]{4em}{\includegraphics[scale=0.25]{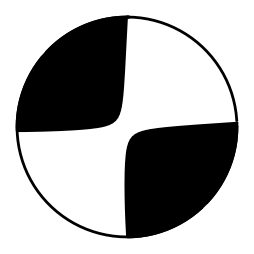}} & 
\parbox[c]{4em}{\includegraphics[scale=0.25]{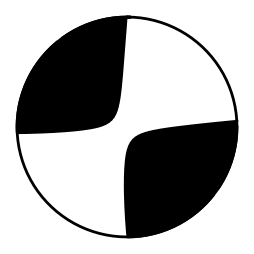}} &
\parbox[c]{4em}{\includegraphics[scale=0.25]{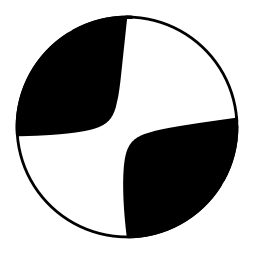}} &
\parbox[c]{4em}{\includegraphics[scale=0.25]{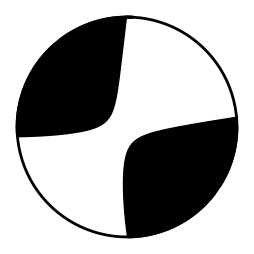}} &
\parbox[c]{4em}{\includegraphics[scale=0.25]{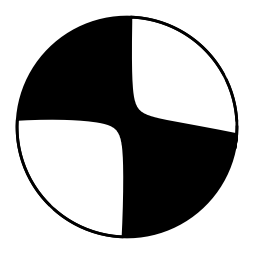}} \\
\tabucline[1pt]{-}
\end{tabu}
\label{tab:ex2cavsheatensimag}
\end{table}

\begin{figure}[H]   
    \center
    \includegraphics[scale=0.6]{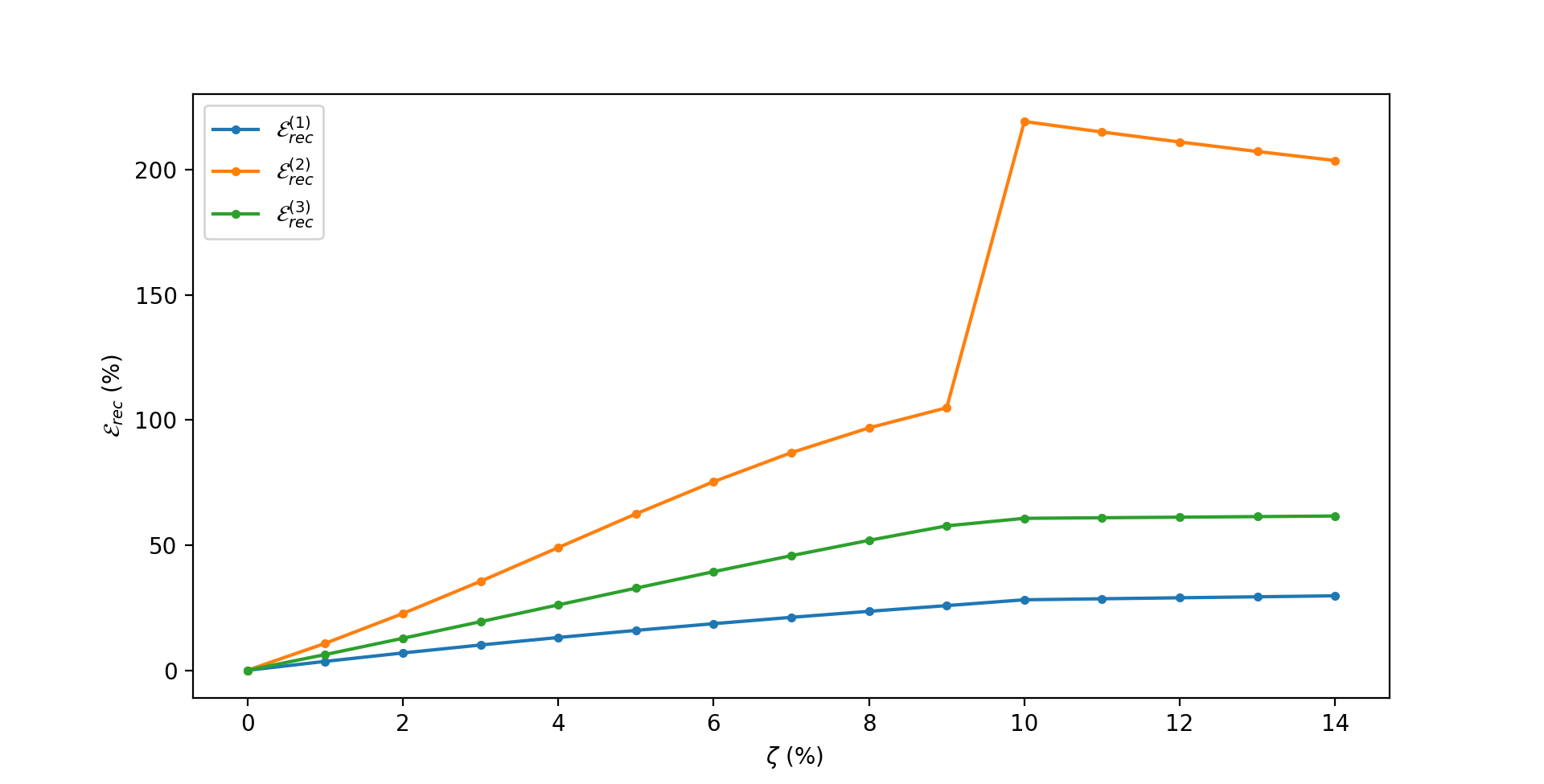}
    \caption{Contemporaneous \emph{on-grid} events: Evolution of $\mathcal{E}^{(n)}\rec$  with the noise level.}
    \label{ex2:norm2}
\end{figure}

The relative error function $\mathcal{E}^{(n)}\rec$ from \eqref{eq:error-rec} with respect to the noise level is presented in Fig.~\ref{ex2:norm2}. From the display, we observe that for the one out of the three events, $\mathcal{E}^{(n)}\rec$ exhibits a sharp jump at the noise level between 9 and 10\%, which coincides with the failure of the algorithm to correctly identify the respective event locations (see Fig.~\ref{fig:example2location}). The reconstructed seismic moment tensors are, however, far more sensitive to noisy data as discussed next.

\subsubsection{Contemporaneous off-grid events $\widehat{\bfx}^{(n)} \notin X$.}

\noindent Next, we consider the reconstruction of three micro-seismic sources representing shear, tensile, and mixed-mode events such that $\widehat{\bfx}^{(n)} \notin X$. In particular, the locations $\widehat{\bfx}^{(n)}$ and intensities $\gamma^{(n)}$ are reported in Table \ref{tab:example2B}.
\begin{table}[H]
    \centering
    \caption{Contemporaneous off-grid micro-seismic sources.}
    \begin{tabu}{cccc} \tabucline[1pt]{-} \addlinespace[2pt]
        $n$th & event & $\gamma^{(n)}$ & $\widehat{\bfx}^{(n)}$ \\ 
         \addlinespace[2pt] \tabucline[1pt]{-} \addlinespace[2pt]
         $1$ & shear      & $(1.0+2.0i)\times 10^{-10}$ & $(0.03808, 0.03200, 0.08533)$ \\
         $2$ & mixed-mode & $(1.0+2.0i)\times 10^{-10}$ & $(0.11733, 0.12267, 0.11467)$ \\  
         $3$ & tensile    & $(2.0+2.0i)\times 10^{-10}$ & $(0.08283, 0.07867, 0.11413)$ \\
         \tabucline[1pt]{-}
    \end{tabu}
    \label{tab:example2B}
\end{table}

\begin{figure}[H]   
    \center
    \subfigure[$4\%$]{\includegraphics[scale=0.5]{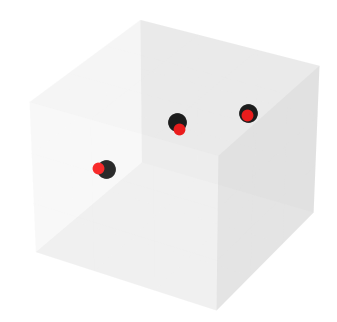}\label{fig:loc4}}\qquad
    \subfigure[$8\%$]{\includegraphics[scale=0.5]{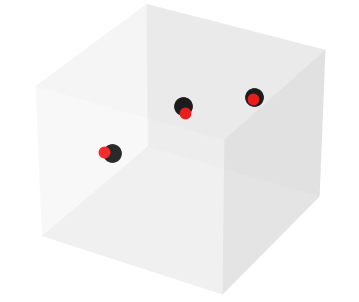}\label{fig:loc8}}\qquad
    \subfigure[$12\%$]{\includegraphics[scale=0.5]{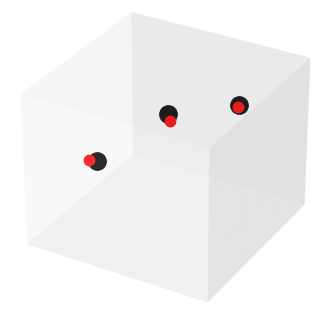}\label{fig:loc12}}\qquad
    \subfigure[$13\%$]{\includegraphics[scale=0.5]{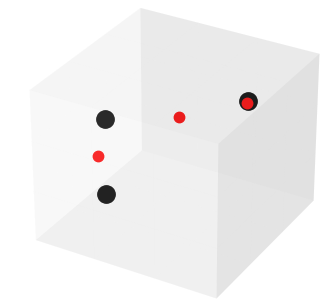}\label{fig:loc13}}
    \caption{Contemporaneous \emph{off-grid} events: Multi-event location results for $4\%$, $8\%$,  $12\%$ and $13\%$ of noise, assuming 9 sensors distributed as in Fig.~\ref{fig:sensor9}.}
    \label{fig:example2Blocation}
\end{figure}

Next, we consider the reconstruction using nine sensors distributed as in Fig.~\ref{fig:sensor9}. The reconstructed fault locations for $4\%$, $8\%$, $12\%$ and $13\%$ are presented in Fig.~\ref{fig:example2Blocation}. The algorithm finds the locations $\bfxi^\star \subset X$ closest to the true ones $\widehat{\bfxi}$ up to $12\%$ of noise. 

\begin{figure}[H]   
    \center
    \includegraphics[scale=0.6]{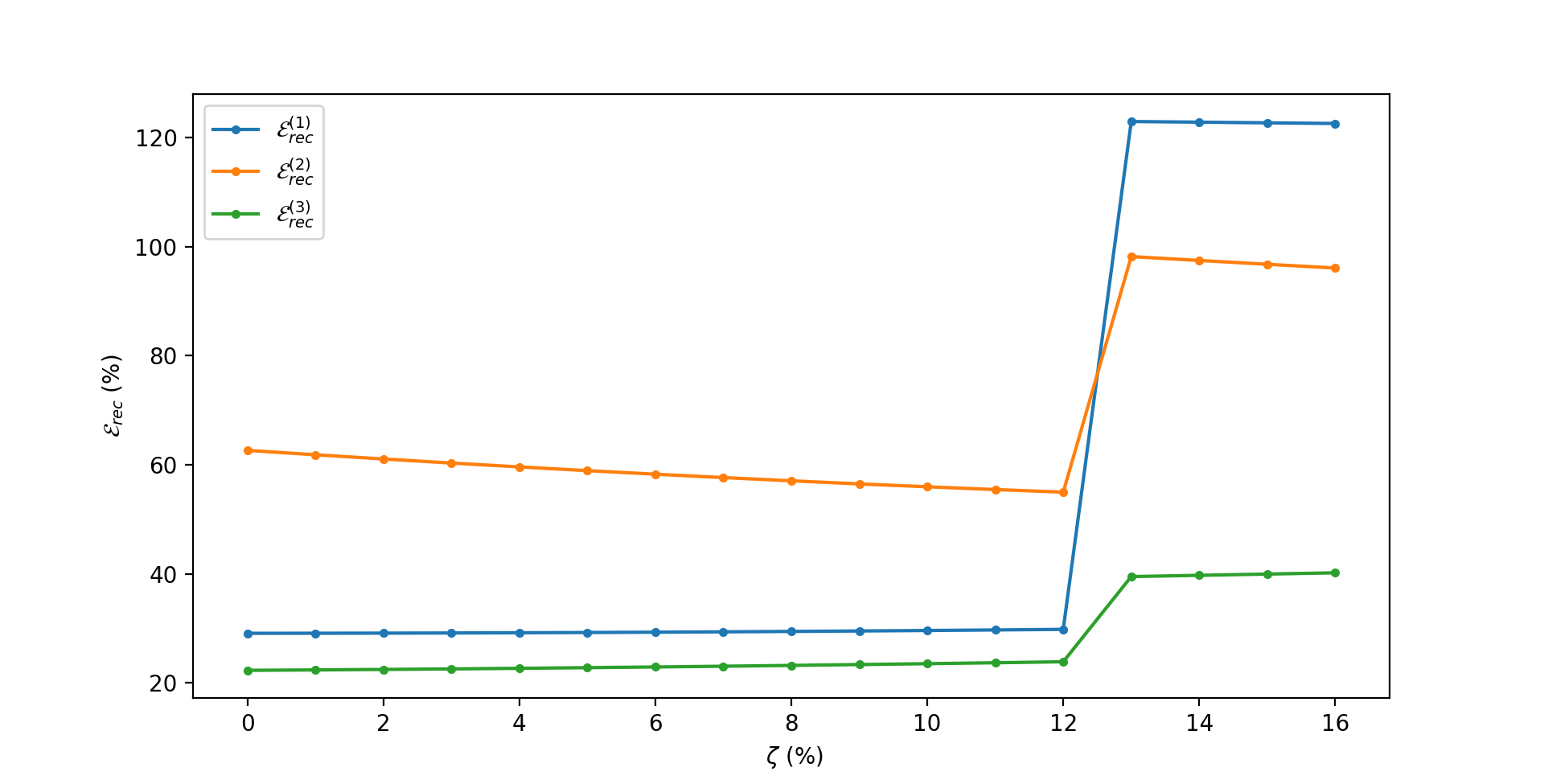}
    \caption{Contemporaneous \emph{off-grid} events: Evolution of $\mathcal{E}^{(n)}\rec$ with the noise level.}
    \label{fig:norm4}
\end{figure}

Evolution of the relative error $\mathcal{E}^{(n)}\rec$ for the noise levels between 8\% and 16\% is shown in Fig.~\ref{fig:norm4}, where we can observe a consequence of the fact that two out of the three event locations are ``lost'' by the reconstruction for $\zeta \geq 13\%$, see Fig.~\ref{fig:example2Blocation}. The focal mechanisms representing the moment tensor associated with the shear, tensile and mixed-mode events are shown in Tables \ref{tab:shear-tensile} and \ref{tab:ex3B-imaginary}  in terms of the (in-phase) real and (out-of-phase) imaginary components, respectively.  In these tables, the first column presents the results for shear, the middle column for tensile, and the last column for mixed-mode events.
\begin{table}[H]
\center
\caption{Contemporaneous \emph{off-grid} events: real component of the reconstructed source mechanisms for the shear, tensile, and mixed-mode events versus the noise level.}
\begin{tabu}{cccccccc} \tabucline[1pt]{-}
target  & $\zeta = 0\%$ & $\zeta = 2\%$ & $\zeta = 4\%$ & $\zeta = 6\%$ & $\zeta = 10\%$& $\zeta = 12\%$& $\zeta = 13\%$ \\ 
\tabucline[1pt]{-}
\parbox[c]{4em}{\includegraphics[scale=0.25]{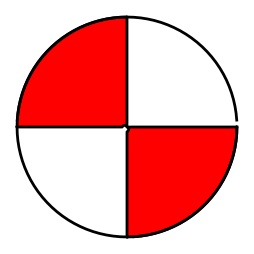}}  & 
\parbox[c]{4em}{\includegraphics[scale=0.25]{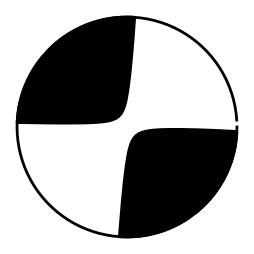}} & 
\parbox[c]{4em}{\includegraphics[scale=0.25]{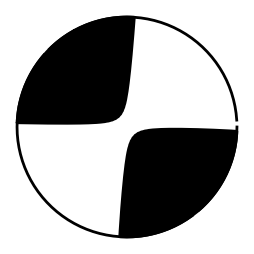}} & 
\parbox[c]{4em}{\includegraphics[scale=0.25]{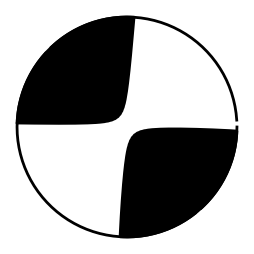}} &
\parbox[c]{4em}{\includegraphics[scale=0.25]{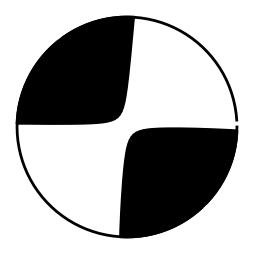}} &
\parbox[c]{4em}{\includegraphics[scale=0.25]{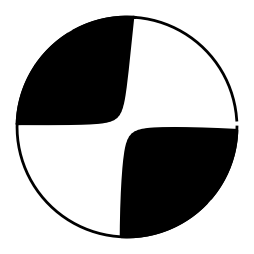}} &
\parbox[c]{4em}{\includegraphics[scale=0.25]{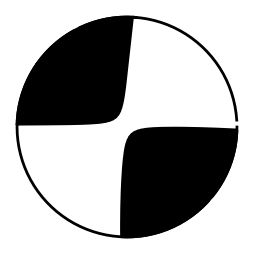}}&
\parbox[c]{4em}{\includegraphics[scale=0.25]{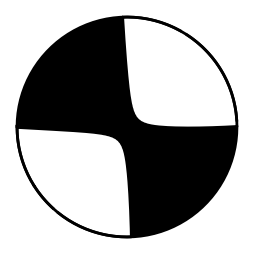}}\\
\parbox[c]{4em}{\includegraphics[scale=0.25]{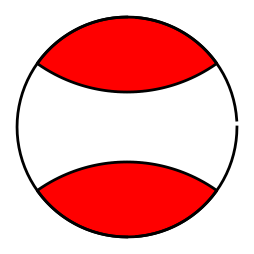}}   & 
\parbox[c]{4em}{\includegraphics[scale=0.25]{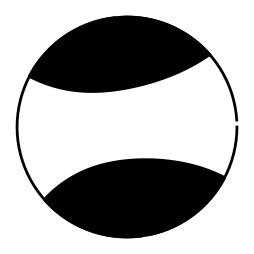}} & 
\parbox[c]{4em}{\includegraphics[scale=0.25]{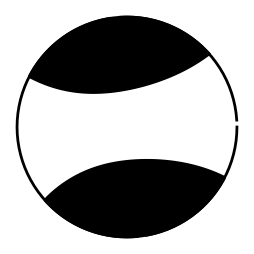}} & 
\parbox[c]{4em}{\includegraphics[scale=0.25]{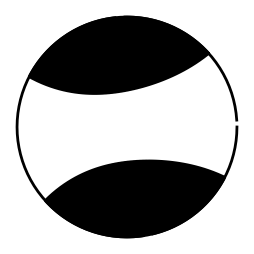}} &
\parbox[c]{4em}{\includegraphics[scale=0.25]{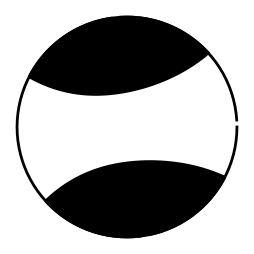}} &
\parbox[c]{4em}{\includegraphics[scale=0.25]{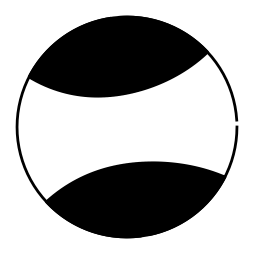}} &
\parbox[c]{4em}{\includegraphics[scale=0.25]{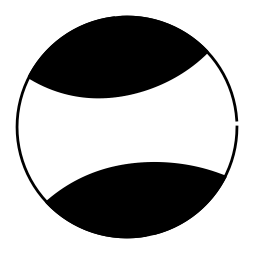}}&
\parbox[c]{4em}{\includegraphics[scale=0.25]{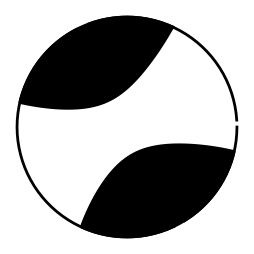}}\\
\parbox[c]{4em}{\includegraphics[scale=0.25]{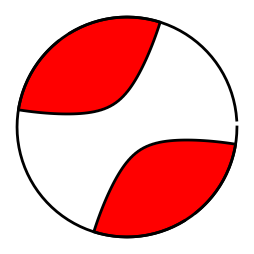}}  & 
\parbox[c]{4em}{\includegraphics[scale=0.25]{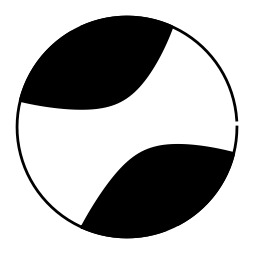}} & 
\parbox[c]{4em}{\includegraphics[scale=0.25]{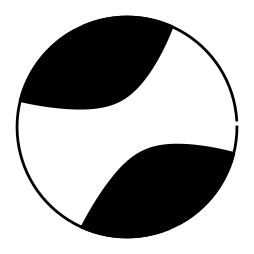}} & 
\parbox[c]{4em}{\includegraphics[scale=0.25]{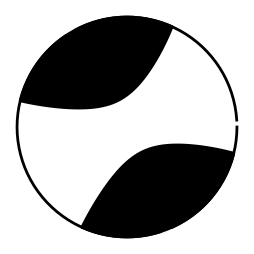}} &
\parbox[c]{4em}{\includegraphics[scale=0.25]{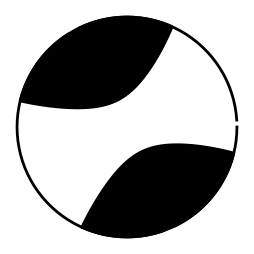}} &
\parbox[c]{4em}{\includegraphics[scale=0.25]{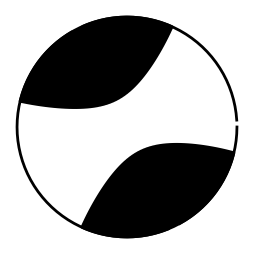}} &
\parbox[c]{4em}{\includegraphics[scale=0.25]{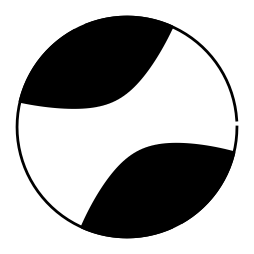}}&
\parbox[c]{4em}{\includegraphics[scale=0.25]{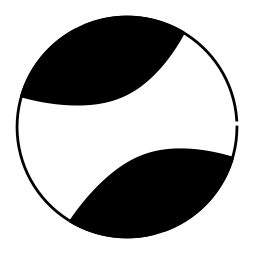}}\\
\tabucline[1pt]{-}
\end{tabu}
\label{tab:shear-tensile}
\end{table}
\begin{table}[H]
\center
\caption{Contemporaneous \emph{off-grid} events: imaginary component of the reconstructed source mechanisms for the shear, tensile, and mixed-mode events versus the noise level.}
\begin{tabu}{cccccccc} \tabucline[1pt]{-}
target  & $\zeta = 0\%$ & $\zeta = 2\%$ & $\zeta = 4\%$  & $\zeta = 6\%$& $\zeta = 10\%$ & $\zeta = 12\%$ & $\zeta = 13\%$ \\ 
\tabucline[1pt]{-}
\parbox[c]{4em}{\includegraphics[scale=0.25]{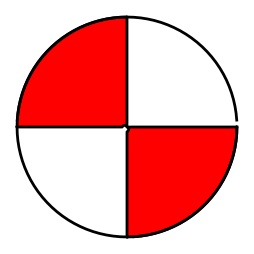}}   & 
\parbox[c]{4em}{\includegraphics[scale=0.25]{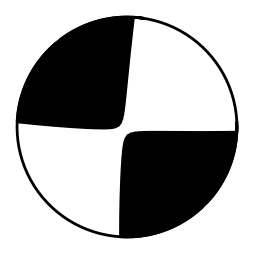}} & 
\parbox[c]{4em}{\includegraphics[scale=0.25]{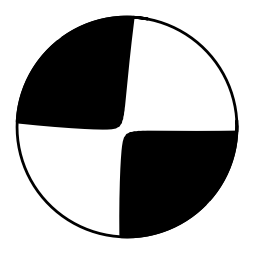}} & 
\parbox[c]{4em}{\includegraphics[scale=0.25]{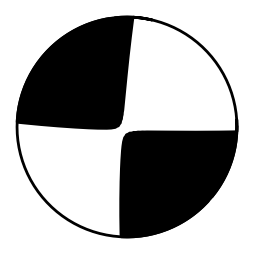}} &
\parbox[c]{4em}{\includegraphics[scale=0.25]{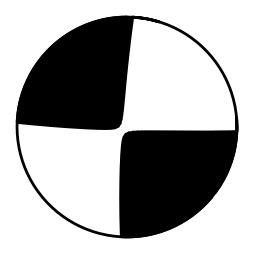}} &
\parbox[c]{4em}{\includegraphics[scale=0.25]{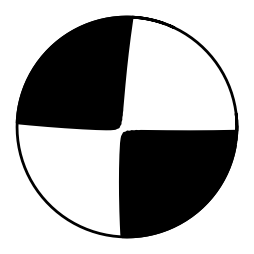}} &
\parbox[c]{4em}{\includegraphics[scale=0.25]{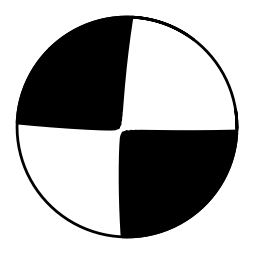}}&
\parbox[c]{4em}{\includegraphics[scale=0.25]{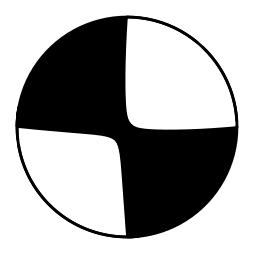}} \\
\parbox[c]{4em}{\includegraphics[scale=0.25]{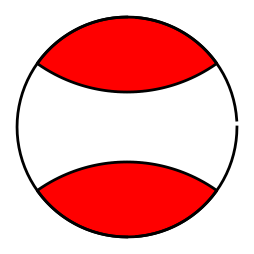}}   & 
\parbox[c]{4em}{\includegraphics[scale=0.25]{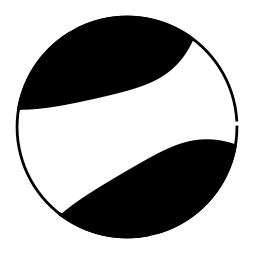}} & 
\parbox[c]{4em}{\includegraphics[scale=0.25]{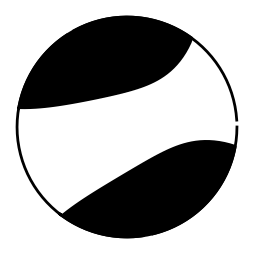}} & 
\parbox[c]{4em}{\includegraphics[scale=0.25]{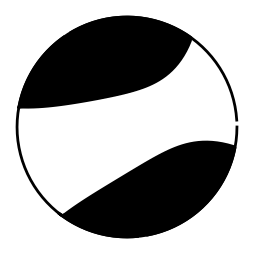}} &
\parbox[c]{4em}{\includegraphics[scale=0.25]{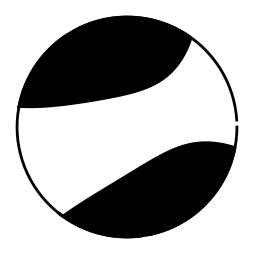}} &
\parbox[c]{4em}{\includegraphics[scale=0.25]{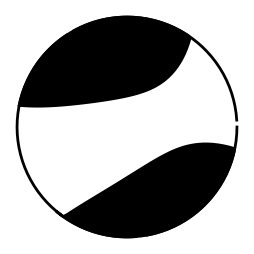}} &
\parbox[c]{4em}{\includegraphics[scale=0.25]{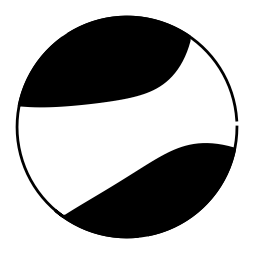}}&
\parbox[c]{4em}{\includegraphics[scale=0.25]{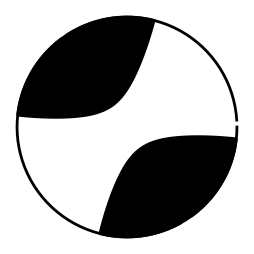}} \\
\parbox[c]{4em}{\includegraphics[scale=0.25]{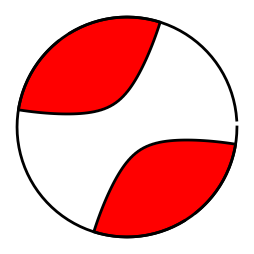}}  & 
\parbox[c]{4em}{\includegraphics[scale=0.25]{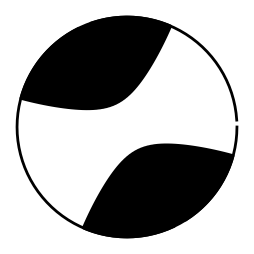}} & 
\parbox[c]{4em}{\includegraphics[scale=0.25]{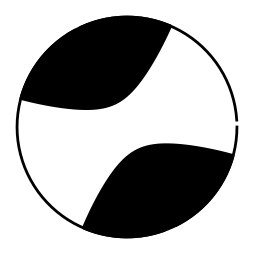}} & 
\parbox[c]{4em}{\includegraphics[scale=0.25]{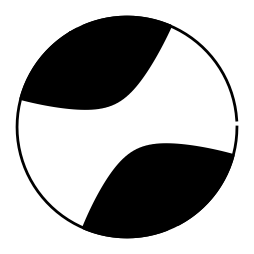}} &
\parbox[c]{4em}{\includegraphics[scale=0.25]{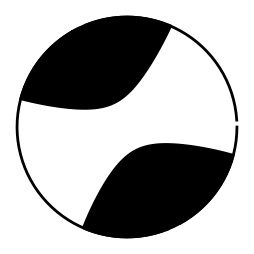}} &
\parbox[c]{4em}{\includegraphics[scale=0.25]{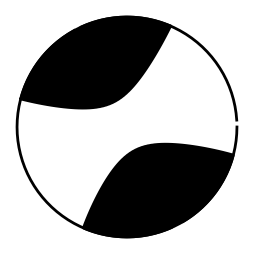}} &
\parbox[c]{4em}{\includegraphics[scale=0.25]{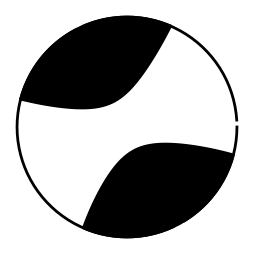}}&
\parbox[c]{4em}{\includegraphics[scale=0.25]{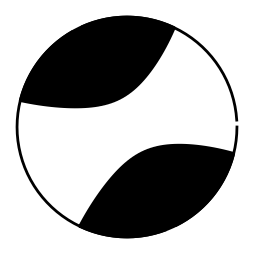}}\\
\tabucline[1pt]{-}
\end{tabu}
\label{tab:ex3B-imaginary}
\end{table}

\subsection{Analysis of transient events}

\noindent For a deeper insight into the performance of the source reconstruction scheme, we next consider \emph{transient} micro-seismic events according to 
\begin{equation} 
\bff(\bfx,t) = \sum_{n=1}^{N} \boldsymbol{\mathcal{M}}^{(n)}(t) \sip \nabla_{\!\!\bfx^{(n)}} \delta(\bfx-\bfx^{(n)}),
\label{eq:transient}
\end{equation}
where
\begin{equation} 
\boldsymbol{\mathcal{M}}^{(n)}(t) =  \bfM^{(n)} \, S^{(n)}(t);
\label{eq:timeMoment}
\end{equation}
$\bfM^{(n)}\!\!\in\! \mathds{R}^{3 \times 3}$ is constant moment tensor, and $S^{(n)}(t)$ is a scalar time trace given by the Morlet wavelet
\begin{equation}
S^{(n)}(t) = a_n e^{-(t/\varsigma_n)^2} \hh \cos(\omega^{(n)} t) 
\label{morspec}
\end{equation}
that is characterized by the carrier i.e.~center frequency~$\omega^{(n)}$, amplitude $a_n$, and ``width'' $\varsigma_n$ which controls its temporal localization. In general, $\boldsymbol{\mathcal{M}}^{(n)}(t)$ may change its tensor character during the event and so may not be amenable to decomposition~\eqref{eq:timeMoment}; for the remainder of this work, however, we retain this simplifying assumption. To handle~\eqref{eq:transient}--\eqref{morspec} by the frequency-domain reconstruction algorithm, we consider the Fourier transform pair 
\begin{equation}
\breve{S}^{(n)}(\omega) = \tfrac{1}{\sqrt{2\pi}}\int_{-\infty}^{+\infty} S^{(n)}(t) \, e^{i \omega t} \dd t, \qquad 
S^{(n)}(t) = \tfrac{1}{\sqrt{2\pi}}\int_{-\infty}^{+\infty} \breve{S}^{(n)}(\omega) \, e^{-i \omega t} \dd \omega.
\end{equation}
To handle contemporaneous events with different onset times, we consider the shifted time traces
\[
S^{(n)}_{\tau_n}(t) \,:=\, S^{(n)}(t-\tau_n), \quad n=\overline{1,N}
\]
with $\tau_n$ being a time ``delay'' of the~$n$th event. In this setting, the Fourier transforms of the respective moment tensors can be written as 
\begin{equation} 
\breve{\!\!\boldsymbol{\mathcal{M}}}^{(n)}_{\tau_n}(\omega) = \bfM^{(n)} \, \breve{S}^{(n)}_{\tau_n}(\omega), \qquad n=\overline{1,N}
\label{eq:harmonicMoment}
\end{equation}
where 
\begin{equation} 
\breve{S}^{(n)}_{\tau_n}(\omega) = \breve{S}^{(n)}(\omega) \, e^{i \omega \tau_n}, \qquad 
\breve{S}^{(n)}(\omega) =
\tfrac{\sqrt{2}}{4} a_n \hh\varsigma_n\hh 
e^{-\frac{1}{4}\varsigma_n^2 \hh (\omega^{(n)}+\omega)^2}
\big(1+e^{\varsigma_n^2\hh\omega^{(n)}\hh\omega}\big),
\label{Sn-Four}
\end{equation}
with~$\breve{S}^{(n)}$ describing a Gaussian distribution centered at $\omega=\omega^{(n)}$.  

\begin{figure}[H]   
    \center
    \subfigure[]{\includegraphics[scale=0.75]{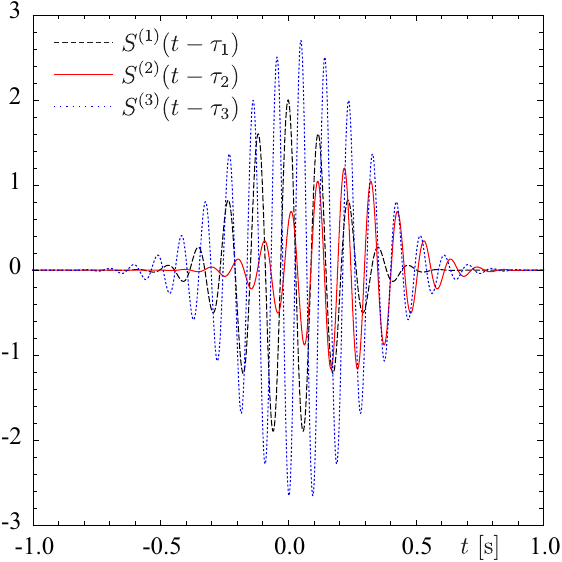}\label{fig:example3:signal:timeDomain}}\qquad
    \subfigure[]{\includegraphics[scale=0.75]{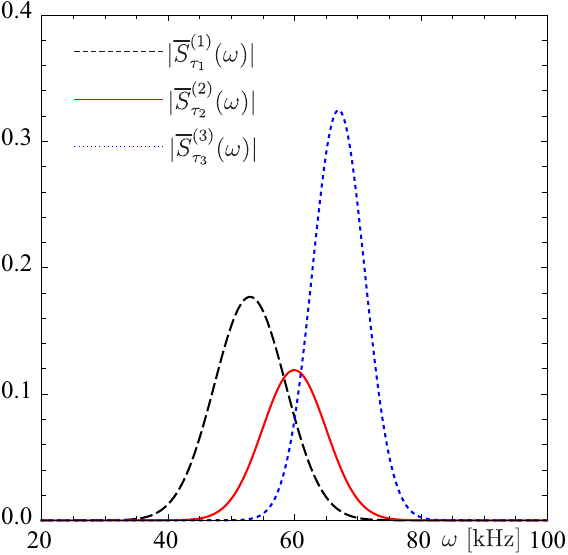} \label{fig:example3:signal:frequencyDomain}}
    \caption{Wavelets used to generate the fault signals: (a) $S^{(n)}_{\tau_n}(t)$ in the time domain, and (b) $|\overline{S}^{(n)}_{\tau_n}(\omega)|$ in the frequency domain ($n=\overline{1,3}$).}
    \label{fig:signal}
\end{figure}

On the basis of \eqref{eq:transient}--\eqref{eq:harmonicMoment}, our objective consists in reconstructing the multi-frequency source distribution 
\begin{equation} 
\boldsymbol{f} (\bfx;\omega_j) =  
\sum_{n=1}^{{N}} \breve{S}^{(n)}_{\tau_n}(\omega_j) \, 
\bfM^{(n)}\sip \nabla_{\!\!\bfx^{(n)}} \delta(\bfx - {\bfx}^{(n)}), \qquad j=\overline{1,J}
\label{eq:harmonicSource}
\end{equation}
where (i) $\bfM^{(n)}$ is assumed to be frequency-independent according to~\eqref{eq:timeMoment}, and (ii) the sampling frequencies~$\omega_j$ are selected to span the spectral content of the (anticipated) micro-seismic events.
Assuming \eqref{eq:harmonicSource} as the source of \eqref{eq:strongZ} evaluated at $\omega=\omega_j$, we denote by $\bfuobs_j$ ($j=\overline{1,J}$) the respective sensory data for the multi-frequency inverse problem. In this vein, we introduce the aggregate cost functional
\begin{equation}
\mathcal{J}(\bfu_1,\cdots,\bfu_J) = 
\tfrac{1}{2} \sum_{j=1}^{J}\int_{\Gamobs} \! \mathfrak{s}(\bfx) \, \|\bfu_j-\bfuobs_j\|^2 \dd\Gamma,
\end{equation}
where $\bfu_j$ is the time-harmonic solution of \eqref{eq:strongOri} due to $\boldsymbol{f}=\boldsymbol{f}(\bfx;\omega_j)$ and $\bfuobs_j$ are the sensory data in  \eqref{eq:strongZ} for computing $\widehat{\boldsymbol{f}}= \widehat{\boldsymbol{f}}(\bfx;\omega_j)$.

In the ensuing numerical example, we consider three contemporaneous events whose intensities $\gamma^{(n)}$, locations $\widehat{\bfx}^{(n)}$, and wavelet parameters ($\omega_n$, $\varsigma_n$, $a_n$ and $\tau_n$) are reported in Table \ref{tab:example3}. We again consider an elastic block with $\ell\!=\!0.16$m, $E\!=\!20$GPa, $\rho\!=\!2300$kg/m$^3$ and $\nu=0.2$. 
 
\begin{table}[H]
    \centering
    \caption{Characteristics of the  transient events.}
    \begin{tabu}{ccccccc} \tabucline[1pt]{-} \addlinespace[2pt]
         event & $\omega_n$ & $\varsigma_n$ & $a_n$ & $\tau_n$ & $\gamma^{(n)}$ &$\widehat{\bfx}^{(n)}$ 
         \\ \addlinespace[2pt] \tabucline[1pt]{-} \addlinespace[2pt]
          1: shear      & $53$kHz & $0.25$ & $2.0$ & $0.00$s & $1.0 \times 10^{-10}$ & $(0.1,0.06,0.02)$ \\  
         2: mixed-mode & $60$kHz & $0.28$ & $1.2$ & $0.22$s & $1.0 \times 10^{-10}$ & $(0.1,0.08,0.02)$ \\         
         3: tensile    & $67$kHz & $0.34$ & $2.7$ & $0.05$s & $1.0 \times 10^{-10}$ & $(0.1,0.08,0.04)$ \\
         \tabucline[1pt]{-}
    \end{tabu}
    \label{tab:example3}
\end{table}

\begin{figure}[H]   
    \center
    \subfigure[]{\includegraphics[scale=0.52]{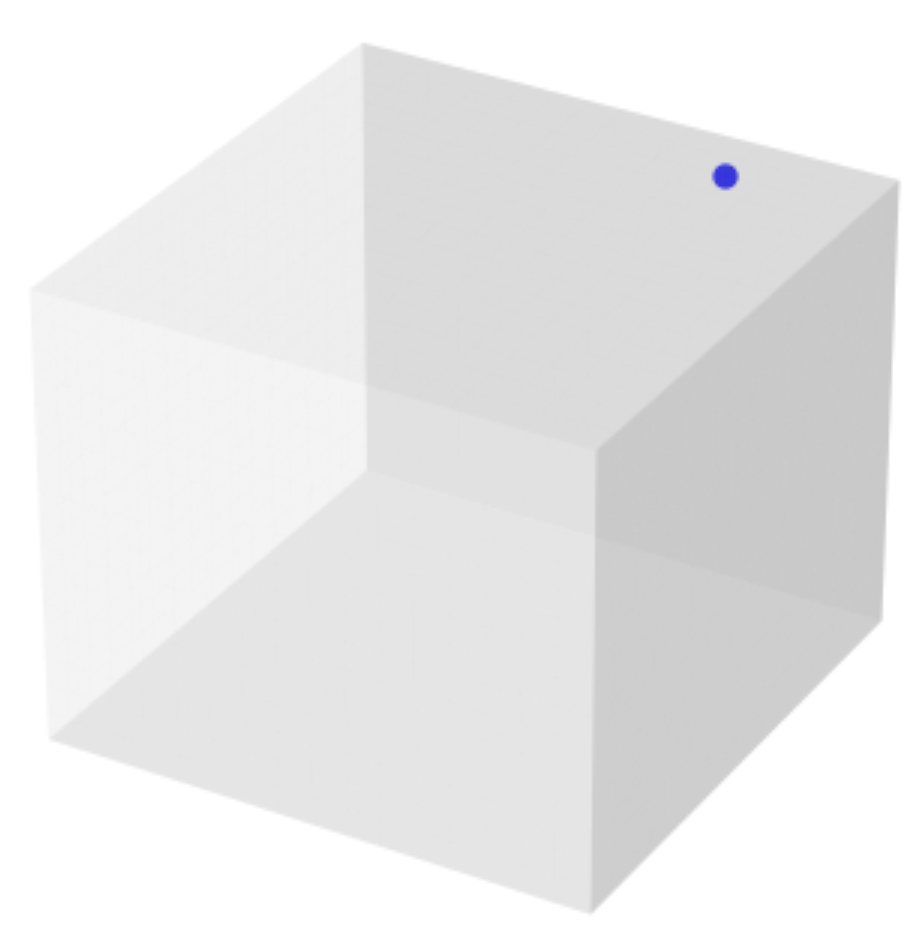}\label{fig:example3:sensor}}\qquad
    \subfigure[]{\includegraphics[scale=0.52]{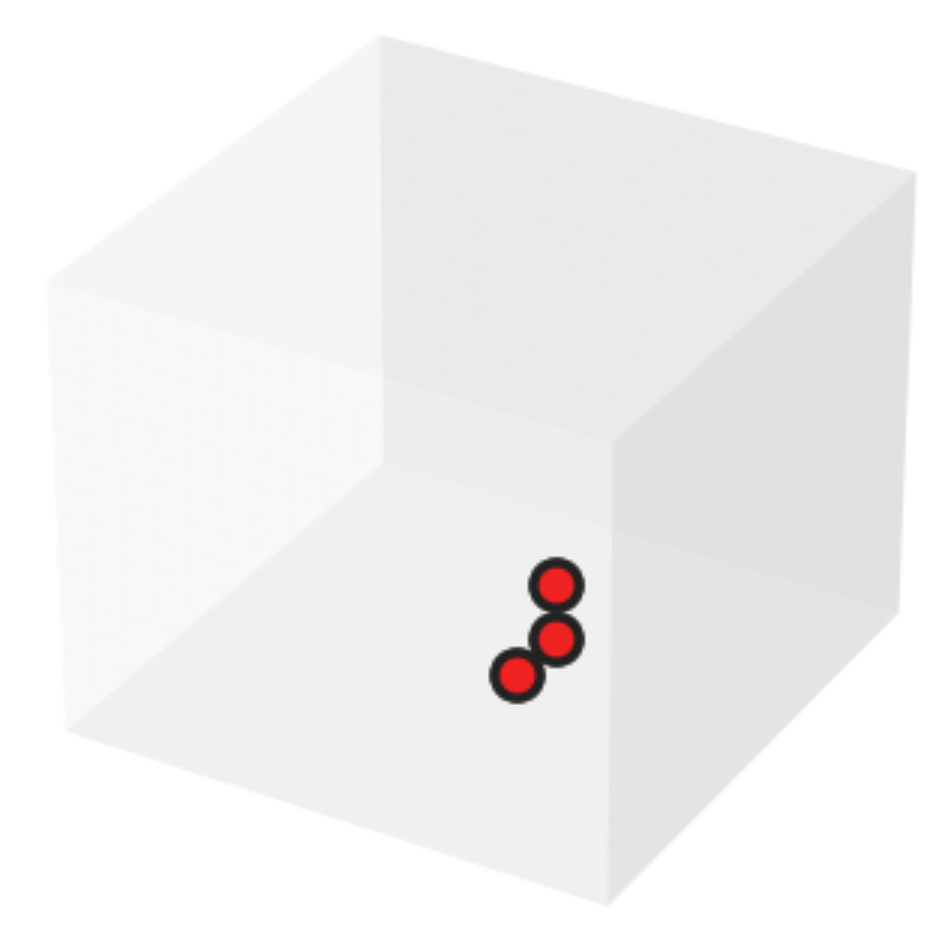}\label{fig:example3:uniaxial:recovered}} \qquad
    \subfigure[]{\includegraphics[scale=0.52]{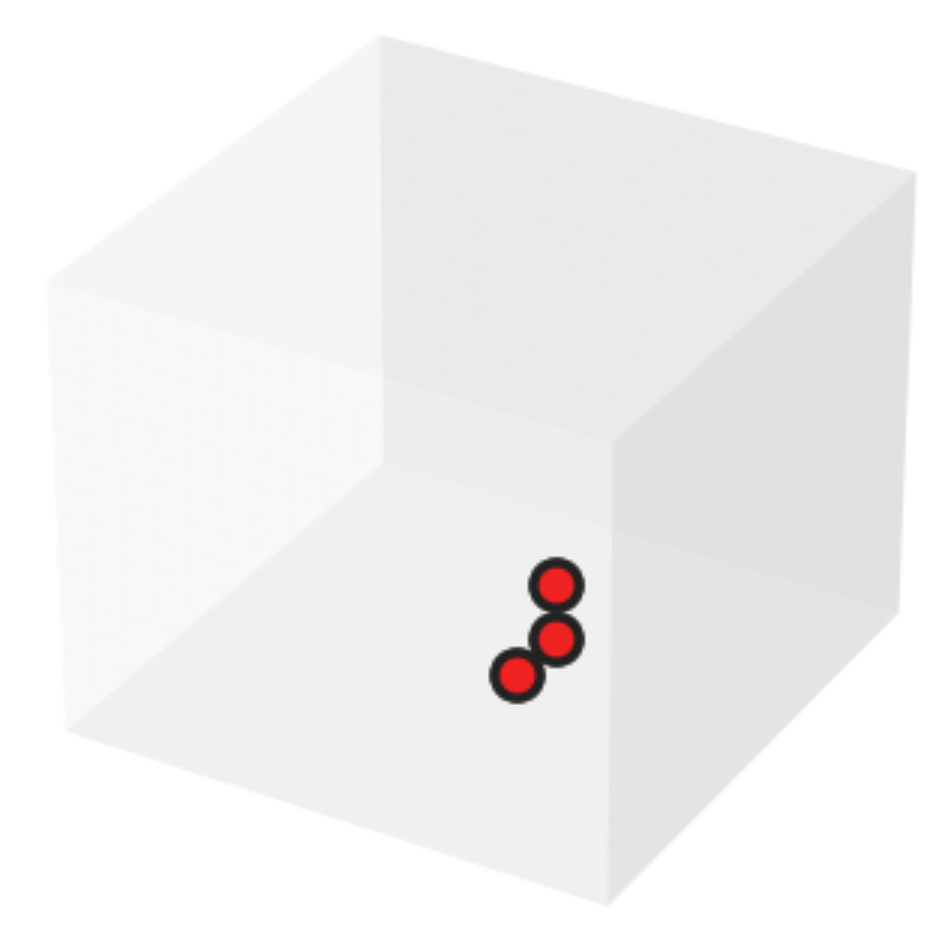}\label{fig:example3:recovered}}
    \caption{Simultaneous reconstruction of three on-grid transient events, noise-free data: (a) the placement of a single accelerometer used to capture the data; (b) event locations for setup I which assumes uniaxial accelerometer and 11 observation frequencies, and (c) event locations for setup II which assumes triaxial accelerometer and 5 observation frequencies (red circles: true locations, black circles: reconstructed locations).}
\end{figure}

In the experiment, we attempt to reconstruct the three events with a \emph{single accelerometer} attached to the top surface of the specimen as shown in Fig.~\ref{fig:example3:sensor}. Two sensing scenarios are considered: setup I deploys a uniaxial motion sensor (capturing the acceleration normal to the boundary), while setup II assumes availability of a triaxial accelerometer. In setup I, the observation frequencies $\omega_j$ ($j\!=\!\overline{1,11}$) span the range [40,80] kHz with $4$kHz separation, while in setup II we cover the same frequency range with $10$kHz separation ($\omega_j$, $j\!=\!\overline{1,5}$).

Fig.~\ref{fig:example3:uniaxial:recovered} and Fig.~\ref{fig:example3:recovered} show respectively the event locations for setup I and setup II, where the red (resp.~black) circles represent the exact (resp. reconstructed) locations. Note that the mutual proximity of micro-seismic events generally creates difficulties in their reconstruction; despite such an impediment, the algorithm is successful in reconstructing all three events using 11 (resp.~5) probing frequencies captured by a single uniaxial (resp. triaxial) sensor. Here it is useful to note that due to multiple scattering by the domain boundaries, the use of different frequencies (and so wavelengths) effectively amounts to having different ``vantage points'' from which the micro-seismic events are being observed. In this vein, the drawbacks of having a uniaxial sensor in setup I are compensated by the use of a denser set of observation frequencies. Such compensation is facilitated by the premise that (for each event) the seismic moment tensor is time- and thus frequency-invariant, see~\eqref{eq:timeMoment} and~\eqref{eq:harmonicMoment}, which then allows different frequencies to ``see'' the same event (in terms of both location and seismic moment tensor) and so reinforce each other in the inversion process.

Table \ref{tab:example3:beachball} lists the exact and reconstructed moment tensors for both setups, described in terms of focal mechanisms. Again, for both setups and all three events, the reconstructed source mechanism is remarkably close to the true signature.
\begin{table}[H]
\center
\caption{Simultaneous reconstruction of three on-grid transient events, noise-free data: exact and reconstructed seismic moment tensors, represented as focal mechanisms}
\begin{tabu}{cccc} \tabucline[1pt]{-} \addlinespace[2pt]
Event &  target \qquad & \begin{tabu}{c}reconstruction \\ (setup I) \end{tabu} & \!\!\!\begin{tabu}{c}reconstruction \\ (setup II) \end{tabu} \\ \addlinespace[2pt]
\tabucline[1pt]{-}
1: shear &  
\parbox[c]{4em}{\includegraphics[scale=0.25]{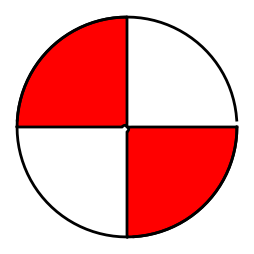}}  & \quad
\parbox[c]{4em}{\includegraphics[scale=0.25]{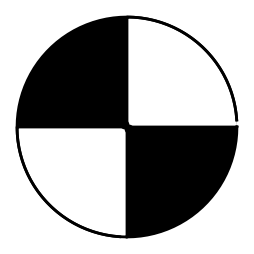}} &
\parbox[c]{4em}{\includegraphics[scale=0.25]{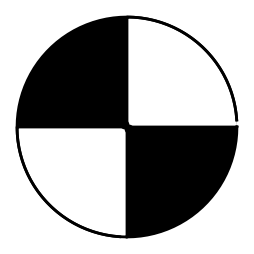}} \\
2: mixed-mode &  
\parbox[c]{4em}{\includegraphics[scale=0.25]{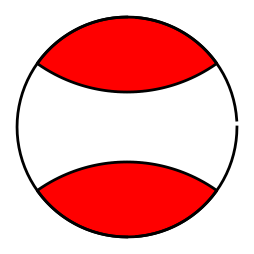}}   & \quad
\parbox[c]{4em}{\includegraphics[scale=0.25]{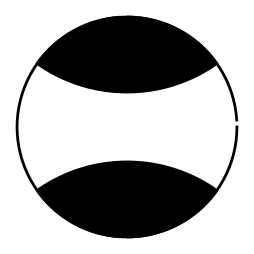}} &
\parbox[c]{4em}{\includegraphics[scale=0.25]{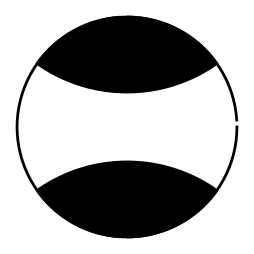}} \\
3: tensile &  
\parbox[c]{4em}{\includegraphics[scale=0.25]{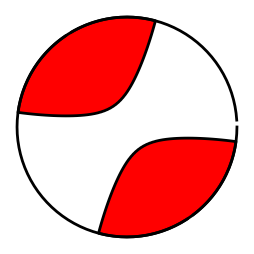}}  & \quad
\parbox[c]{4em}{\includegraphics[scale=0.25]{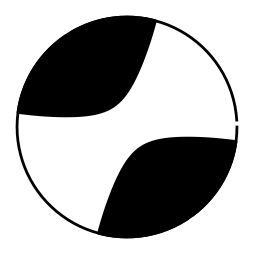}} &
\parbox[c]{4em}{\includegraphics[scale=0.25]{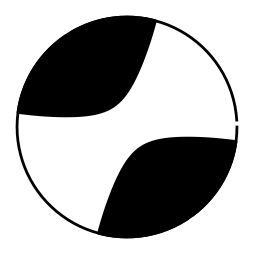}} \\
\tabucline[1pt]{-}
\end{tabu}
\label{tab:example3:beachball}
\end{table}

\section{Concluding Remarks} \label{sec:conclusions}

 \noindent In this study we propose an algorithm for the spatial reconstruction and characterization of micro-seismic events in a three-dimensional bounded elastic body, whose geometry and material properties are known beforehand, via joint source location and moment tensor inversion. Using selected frequency component(s) of the “acoustic emission” traces at sensor locations, the inverse solution is established on the basis of (i) time-harmonic (forward) elastodynamic model and (ii) the concept of topological derivative. On exploiting an equivalence between the elastic wavefield generated by the creation of a new micro-surface and that stemming from a suitable set of dipoles and double-couples, we formulate the inverse problem as that for the real (in-phase) and imaginary (out-of-phase) components of the seismic moment tensor at trial grid locations. In this way, the optimal solution is obtained via a combinatorial search over a prescribed grid -- which caters for successive refinements over the region(s) of interest. The analysis is illustrated by numerical experiments highlighting the key features of the inversion scheme including the reconstruction of multiple (i.e.~contemporaneous) events,  localization of the ``off-grid'' micro-seismic events, and the ability to handle noisy data. The results in particular highlight the utility of multi-frequency event reconstruction toward reducing the demand on the number of sensing locations. 

\section*{Acknowledgements}

\noindent This research was supported in part by CNPq (Brazilian Research Council), CAPES (Brazilian Higher Education Staff Training Agency), FAPERJ (Research Foundation of the State of Rio de Janeiro), and the \emph{Center on Geo-processes in Mineral Carbon Storage} -- an Energy Frontier Research Center funded by the U.S. Department of Energy, Office of Science, Basic Energy Sciences at the University of Minnesota under award \# DE-SC0023429. 

\bibliographystyle{plainnat}
\bibliography{BibLNCC}

\begin{thebibliography}{29}
\providecommand{\natexlab}[1]{#1}
\providecommand{\url}[1]{\texttt{#1}}
\expandafter\ifx\csname urlstyle\endcsname\relax
  \providecommand{\doi}[1]{doi: #1}\else
  \providecommand{\doi}{doi: \begingroup \urlstyle{rm}\Url}\fi

\bibitem[Aki and Richards(2009)]{AkiBook2009}
K~Aki and P.G. Richards.
\newblock \emph{Quantitative Seismology}.
\newblock University Science Books, Sausalito, California, 2009.

\bibitem[Amad et~al.(2020)Amad, Novotny, and Guzina]{AmadIJSS2020}
AAS Amad, AA~Novotny, and BB~Guzina.
\newblock On the full-waveform inversion of seismic moment tensors.
\newblock \emph{International Journal of Solids and Structures}, 202:\penalty0
  717--728, 2020.

\bibitem[Ammari and Kang(2004)]{AmmariBook2004}
H~Ammari and H~Kang.
\newblock \emph{Reconstruction of small inhomogeneities from boundary
  measurements}.
\newblock Lectures Notes in Mathematics vol. 1846. Springer-Verlag, Berlin,
  2004.

\bibitem[Ammari et~al.(2013)Ammari, Garnier, Jing, Kang, Lim, S{\o}lna, and
  Wang]{AmmariBook2013}
H~Ammari, J~Garnier, W~Jing, H~Kang, M~Lim, K~S{\o}lna, and H~Wang.
\newblock \emph{Mathematical and statistical methods for multistatic imaging},
  volume 2098.
\newblock Springer, Switzerland, 2013.

\bibitem[Carpinteri et~al.(2012)Carpinteri, Xu, Lacidogna, and
  Manuello]{carpinteri2012reliable}
A~Carpinteri, J~Xu, G~Lacidogna, and A~Manuello.
\newblock Reliable onset time determination and source location of acoustic
  emissions in concrete structures.
\newblock \emph{Cement and Concrete Composites}, 34:\penalty0 529--537, 2012.

\bibitem[{Dalla Riva} et~al.(2021){Dalla Riva}, {de Cristoforis}, and
  Musolino]{RivaBook2021}
M~{Dalla Riva}, ML~{de Cristoforis}, and P~Musolino.
\newblock \emph{Singularly perturbed boundary value problems: A functional
  analytic approach}.
\newblock Springer Nature Switzerland, 2021.

\bibitem[Elhadidy et~al.(2023)Elhadidy, Alarab, Salam, and
  Faried]{ElhadidyJAG2023}
M~Elhadidy, M.E. Alarab, M~Salam, and A.~Faried.
\newblock Source parameters and moment tensor inversion of april 11, 2020
  (mb=4.4), the significant {El-Negalah} earthquake, {M}atrouh {E}gypt.
\newblock \emph{NRIAG Journal of Astronomy and Geophysics}, 12:\penalty0
  58--71, 2023.

\bibitem[Gilbert(1973)]{GilbertPTRS1973}
F~Gilbert.
\newblock Derivation of source parameters from low-frequency spectra.
\newblock \emph{Philosophical Transactions of the Royal Society of London.
  Series A, Mathematical and Physical Sciences}, 274:\penalty0 369--371, 1973.

\bibitem[Ihlenburg and Babu{\v{s}}ka(1995)]{BabuskaCMA1995}
F~Ihlenburg and I~Babu{\v{s}}ka.
\newblock Finite element solution of the helmholtz equation with high wave
  number part i: The h-version of the fem.
\newblock \emph{Computers MatH Applic.}, 30\penalty0 (9):\penalty0 9--37, 1995.

\bibitem[Koerner et~al.(1981)Koerner, McCabe, and Jr]{KoenerRM1981}
RM~Koerner, WM~McCabe, and AE~Lord Jr.
\newblock Overview of acoustic emission monitoring of rock structures.
\newblock \emph{Rock Mechanics}, 14:\penalty0 27--35, 1981.

\bibitem[Labuz et~al.(2001)Labuz, Cattaneo, and Chen]{labuz2001acoustic}
JF~Labuz, S~Cattaneo, and L-H Chen.
\newblock Acoustic emission at failure in quasi-brittle materials.
\newblock \emph{Construction and Building Materials}, 15:\penalty0 225--233,
  2001.

\bibitem[Machado et~al.(2017)Machado, Angelo, and Novotny]{MachadoM2AS2017}
TJ~Machado, JS~Angelo, and AA~Novotny.
\newblock A new one-shot pointwise source reconstruction method.
\newblock \emph{Mathematical Methods in the Applied Sciences}, 40:\penalty0
  1367--1381, 2017.

\bibitem[Maji et~al.(1990)Maji, Ouyang, and Shah]{maji1990fracture}
AK~Maji, C~Ouyang, and SP~Shah.
\newblock Fracture mechanisms of quasi-brittle materials based on acoustic
  emission.
\newblock \emph{Journal of Materials Research}, 5:\penalty0 206--217, 1990.

\bibitem[Manthei and Plenkers(2018)]{manthei2018review}
G~Manthei and K~Plenkers.
\newblock Review on in situ acoustic emission monitoring in the context of
  structural health monitoring in mines.
\newblock \emph{Applied Sciences}, 8:\penalty0 1595, 2018.

\bibitem[Maz'ya et~al.(2000)Maz'ya, Nazarov, and Plamenevskij]{MazyaBook2000}
VG~Maz'ya, SA~Nazarov, and BA~Plamenevskij.
\newblock \emph{Asymptotic theory of elliptic boundary value problems in
  singularly perturbed domains. {V}ol. {I}}, volume 111 of \emph{Operator
  Theory: Advances and Applications}.
\newblock Birkh\"auser Verlag, Basel, 2000.

\bibitem[Nazarov and Plamenevskij(1994)]{NazarovBook1991}
SA~Nazarov and BA~Plamenevskij.
\newblock \emph{Elliptic problems in domains with piecewise smooth boundaries},
  volume~13 of \emph{de Gruyter Expositions in Mathematics}.
\newblock Walter de Gruyter \& Co., Berlin, 1994.

\bibitem[Novotny and Soko{\l}owski(2013)]{NovotnyBook2013}
AA~Novotny and J~Soko{\l}owski.
\newblock \emph{Topological derivatives in shape optimization}.
\newblock Interaction of Mechanics and Mathematics. Springer-Verlag, Berlin,
  Heidelberg, 2013.

\bibitem[Novotny and Soko{\l}owski(2020)]{NovotnyBook2020}
AA~Novotny and J~Soko{\l}owski.
\newblock \emph{An introduction to the topological derivative method}.
\newblock Springer Briefs in Mathematics. Springer Nature Switzerland, 2020.

\bibitem[Okoli et~al.(2024)Okoli, Kolawole, Akaolisa, Ikoro, and
  Ozotta]{okoli2024alterations}
AE~Okoli, O~Kolawole, CZ~Akaolisa, DO~Ikoro, and O~Ozotta.
\newblock Alterations in petrophysical and mechanical properties due to
  basaltic rock-{CO2} interactions: comprehensive review.
\newblock \emph{Arabian Journal of Geosciences}, 17:\penalty0 1--23, 2024.

\bibitem[Ono(2014)]{ono2014acoustic}
K~Ono.
\newblock \emph{{A}coustic {E}mission}, pages 1209--1229.
\newblock Springer, 2014.

\bibitem[Quin et~al.(2023)Quin, Li, and et~al.]{QinEnergies2023}
Y~Quin, J~Li, and L.~Huang et~al.
\newblock Microseismic monitoring at the {F}arnsworth {CO2-EOR} field.
\newblock \emph{Energies}, 16:\penalty0 4177, 2023.

\bibitem[Rice(1980)]{RiceJNE1980}
JR~Rice.
\newblock Elastic wave emission from damage processes.
\newblock \emph{Journal of Nondestructive Evaluation}, 1:\penalty0 215--224,
  1980.

\bibitem[Schoberl(1997)]{SchoberlCVS1997}
J~Schoberl.
\newblock An advancing front {2D/3D}-mesh generator based on abstract rules.
\newblock \emph{Computing and Visualization in Science}, 1:\penalty0 41--52,
  1997.

\bibitem[Shearer(2009)]{ShearerBook2009}
PM~Shearer.
\newblock \emph{Introduction to Seismology}.
\newblock Cambridge University Press, 2009.

\bibitem[Shengxiang et~al.(2021)Shengxiang, Qin, Xiling, Xibing, Yu, and
  Daolong]{shengxiang2021study}
L~Shengxiang, X~Qin, L~Xiling, L~Xibing, L~Yu, and C~Daolong.
\newblock Study on the acoustic emission characteristics of different rock
  types and its fracture mechanism in brazilian splitting test.
\newblock \emph{Frontiers in Physics}, 9:\penalty0 591651, 2021.

\bibitem[Sjogreen and Petersson(2014)]{SjogreenJSC2014}
B~Sjogreen and NA~Petersson.
\newblock Source estimation by full wave form inversion.
\newblock \emph{J Sci Comput}, 59:\penalty0 247--276, 2014.

\bibitem[Song and Alkhalifah(2019)]{SongJSTAEORS2019}
C~Song and T~Alkhalifah.
\newblock Mircroseismic event estimation on an efficient wavefield inversion.
\newblock \emph{Journal of Selected Topics in Applied Earth Observation and
  Remote Sensing}, 12:\penalty0 4664--4671, 2019.

\bibitem[Song and Toks{\"{o}}z(2011)]{SongG2011}
F~Song and MN~Toks{\"{o}}z.
\newblock Full-waveform based complete moment tensor inversion and source
  parameter estimation from downhole microseismic data for hydrofracture
  monitoring.
\newblock \emph{Geophysics}, 76, 2011.

\bibitem[Stein and Wysession(2003)]{SteinBook2003}
S~Stein and M~Wysession.
\newblock \emph{An Introduction to Seismology, Earthquakes, and Earth
  Structure}.
\newblock Blackwell, 2003.

\end{thebibliography}

\end{document}